\documentclass[12pt,preprint]{aastex} 
\shorttitle{Chandra, HST and Radio Constraints on 3C~280 and 3C~254} 
\shortauthors{Donahue, Daly, \& Horner}

\newcommand{\flux}{{\rm erg \, s^{-1} \, cm^{-2}}} 
\newcommand{\flambda}{{\rm erg \, s^{-1} \, cm^{-2} \, \AA^{-1}}}

\newcommand{\lum}{{\rm erg \, s^{-1}} } 
 
\newcommand{\ltsim}{{\>\rlap{\raise2pt\hbox{$<$}}\lower3pt\hbox{$\sim$}\> 
 
}} 
\newcommand{\gtsim}{{\>\rlap{\raise2pt\hbox{$>$}}\lower3pt\hbox{$\sim$}\> 
 
}}

\newcommand{\mdot}{{\rm M_\odot \, yr^{-1}}}

\received{2002 August 22} 
\begin{document} 
\title{Constraints on the Cluster Environments and 
Hot Spot Magnetic Field Strengths of the Radio Sources 3C~280 and 
3C~254} 
\author{Megan Donahue} 
\affil{Space Telescope Science Institute, 3700 San Martin 
Drive, 
Baltimore, MD 21218} 
\email{donahue@stsci.edu} 
\author{Ruth A. Daly} 
\affil{Department of Physics, Berks-Lehigh Valley College, 
Pennsylvania State University, Reading, 
PA, 19610} 
\email{rdaly@psu.edu} 
\and 
\author{Donald J. Horner} 
\affil{Space Telescope Science Institute, 3700 San Martin 
Drive, 
Baltimore, MD 21218} 
\email{horner@stsci.edu} 
 
\begin{abstract}

We present new Chandra Observatory observations together with 
archival Hubble Space Telescope and radio observations of 
3C~254, a radio quasar at $z=0.734$, and 3C~280, a radio 
galaxy at $z=0.996$.  
We report the detection of X-ray and possible HST  
optical counterparts to the radio hot spots in 3C~280  
and of an X-ray counterpart 
to the western radio hot spot in 3C~254. 
We present constraints on the presence of 
X-ray clusters and on the magnetic field strengths in and 
around the  
radio hot spots for both targets. 
Both sources were thought to be in clusters of galaxies 
based on reports of significant 
extended emission in 
ROSAT PSPC and HRI images. The exquisite spatial resolution of the 
Chandra Observatory allows us to demonstrate that these sources are 
not in hot, massive clusters. The extended emission seen in ROSAT 
observations is resolved by Chandra into point sources, and is likely to be 
X-ray emission associated with  
the radio hot spots of these sources, with possible 
additional contributions from unrelated point sources. 
The intergalactic medium around these 
sources could be dense, but it is demonstrably not dense and hot.  
We conclude 
that radio sources are not reliable signposts of massive clusters 
at moderately high redshifts.  
We present measurements of the X-ray and optical fluxes 
of source features and discuss what physical processes may give 
rise to them.  X-ray synchrotron emission could explain the radio, optical, 
and X-ray hot spot fluxes in 3C~280; this would require 
continuous acceleration of electrons to high 
Lorentz factors since the synchrotron lifetime 
of relativistic electrons that could produce the 
X-ray emission   
would be of order a human lifetime.  
Synchrotron self Compton (SSC) emission with 
or without inverse Compton (IC) emission 
due to scattering of the cosmic microwave background radiation 
can also explain the X-ray emission 
from the hot spots, though these most likely would require that  
some other physical process be invoked to explain  
the optical emission seen in 3C 280. High spatial resolution radio data with 
broad frequency coverage  
of the radio hot spot regions is needed 
to determine which physical process is responsible for 
the detected X-ray emission, and to provide much tighter 
constraints on  
the magnetic field strength of the hot spot plasma. We summarize our 
current constraints on the magnetic field strengths in  
and around the hot spots of 3C 254 and 3C 280.   
\end{abstract} 
 
% 6 keywords allowed 

\section{Introduction}

Since radio sources are visible from across the universe, understanding 
the connections between their observed properties and their physical 
conditions allows us to probe the environment and physics of galaxies 
billions of years ago. 
Radio sources 
are thought to inhabit 
regions of high galaxy density because of an excess of 
companions (e.g. Pascarelle et al. 1996; Le Fevre et al. 1996) and 
large rotation measures which could arise from hot, magnetized 
cluster gas (Carilli et al. 1997.) Astronomers have speculated that 
the most distant radio sources evolve into massive ellipticals (e.g. 
Pentericci et al. 1999); De Vries et al. (2000) show that 
the galaxy hosts are likely to be giant elliptical galaxies.   
Best et al. (1997ab) 
showed that magnitudes, colors, and profiles of 3CR radio galaxies 
at $z\sim1$ are consistent with those of a 
$M\sim 5 \times 10^{11} \mdot$ elliptical with a 
formation redshift of 3-5. The closest example of such a system is 
Cygnus A. For reference, Cygnus A is a powerful radio galaxy 
inhabiting and interacting with the intracluster medium (ICM) of 
a relatively rich cluster of galaxies with an X-ray bolometric 
luminosity of $L_x \sim  6.2 \times 10^{44} \lum$ (converted 
from a 2-10 keV luminosity of $L_x \sim  3.5 \times 10^{44} \lum$),  
and X-ray temperature $T_x \sim 7.7$ keV (Smith et al. 2002.) 
 
In this work, we observed two powerful FRIIb radio sources. 
An FRIIb source is defined as an edge-brightened, classical double 
radio source with a quite regular (cigar-like) shaped radio bridge 
region, and is also classified as a type 1 FRII source by 
Leahy \& Williams (1984).  FRIIb sources, described in 
some detail by Daly (2002), typically have radio powers at least a factor 
of 10 above the classical FRI-FRII break, and are found almost 
exclusively at 
relatively high-redshift (with the well-known exception, Cygnus A). 
For FRIIb sources, it is thought that the ram pressure 
confinement of the radio lobe might be 
used to estimate the ambient gas density using radio data alone, as was 
done 
by Carilli et al (1991) for Cygnus A, Perley \& Taylor (1991) for 
3C295, Wellman, Daly, \& Wan (1997; WDW97 hereafter) 
for a sample of 22 FRIIb sources, and Guerra, Daly, \& Wan 
(2000) for 6 FRIIb sources.  A comparison of radio 
predicted ambient gas densities with X-ray determined ambient 
gas densities suggests that the magnetic field strength in 
the radio bridges of 3C 295 and Cygnus A are about 1/4  
of the minimum energy value (Perley \& Taylor 1991; Carilli et al. 
1991).

If the ram pressure method of determining ambient gas densities 
were confirmed to be reliable by an X-ray assessment of the 
ambient gas densities, then we might be able to use the radio data 
to ascertain the density of the 
ICM at higher redshifts where the X-ray emission is 
nearly impossible to detect with current technology. Such a method would 
be complementary to Sunyaev-Zel'dovich (SZ) techniques since, 
in general, SZ imaging is more difficult near 
powerful radio sources. 
The properties of the ICM at high redshifts, insofar as they reveal  
the virial mass of the cluster, provide important constraints on 
models of evolution of clusters and of large scale structure, strongly 
constraining $\Omega_{matter}$ 
(e.g. Peebles, Daly, \& Juskiewicz 1989; Donahue et al.  1998, 
1999; Donahue \& Voit 1999; Henry 2000, 1997; Eke et al. 1998; Borgani  
et al. 2002).

In order to test this method of assessing densities, 
we observed two very powerful, extended, classical double radio FRIIb 
sources, 3C~254 and 3C~280. Both of these sources had strong 
indications 
of extent in X-ray images from studies using ROSAT 
(Hardcastle \&  Worrall 1999; Crawford et al.  
1999).  
The excellent spatial resolution of the Chandra X-ray Observatory 
allows us to distinguish between 
the compact emission from a central point source and hot spots from  
extended emission 
from a thermal, extended gaseous halo gravitationally confined by 
the potential associated with a cluster of galaxies. 
Typical clusters of galaxies have ICM luminosities 
of $L_x \sim 10^{43-45} \lum$ and gas temperatures of 
$T_x \sim 2-15$ keV. A hot intracluster medium has been 
detected around at least some distant 
3CR radio galaxies, using Chandra's imaging detectors. 3C294 at 
$z=1.786$ 
was found to have diffuse emission of $\sim 5$ keV and a rest-frame 
$0.3-10$ keV X-ray luminosity of $4.5 \times 10^{44} h_{50}^{-2}$ erg 
s$^{-1}$ (Fabian et al. 2001). 3C295 ($z=0.47$) has a $5$ keV ICM, 
detected 
out to $1 h_{50}^{-1}$ Mpc with Chandra (Allen et al. 2001). 3C220.1 
($z=0.62$) is contained within a $5$ keV ICM detected with Chandra 
with a rest-frame 
X-ray luminosity (0.7-12.0 keV) of $5.6 \times 10^{44} \lum$ (Worrall 
et al. 2001). Despite the detections of clusters around other radio  
sources and  
the early assessments of extent in 3C~280 and 3C~254, in this work  
we find that neither of these 
sources is in a cluster of any 
significance ($L_x  \gtrsim 3 \times 10^{43}$ erg s$^{-1}$.) 
We discuss our X-ray observations, our limits on extended X-ray flux, 
and what inferences may be made regarding the ICM  
surrounding these sources. 
 
However, we also detect significant emission from X-ray counterparts 
of the radio hot spots in both sources.  
In this paper, we describe the two targets of our studies in \S2. We 
outline the details of the Chandra observations and analysis in \S3. 
We briefly discuss our reduction and analysis of HST data in \S4. 
The limits placed on the existence of the X-ray clusters and 
the density profiles of each system are contained in \S5.1 and \S5.2, 
along with a review of the archival   
ROSAT data in \S5.3. In \S6.1, we present useful relations 
for computing the expectations for synchrotron self-Compton (SSC) 
processes in hot spots, and apply them to our observations. In 
\S6.2 we discuss the contribution of synchrotron to the X-ray and possible 
optical emission; in 
\S6.3 we do the same for inverse Compton (IC) scattering of the 
microwave background. In all sections we place constraints on the 
magnetic field. In \S7, we summarize our conclusions regarding the 
physical environments of the radio sources, the 
physical processes causing the hot spots, and the magnetic fields in 
those hot spots.

For the purposes of comparing luminosities and length scales between this 
work and previous X-ray research, 
we assume $H_0=50 h_{50}$ km s$^{-1}$ Mpc$^{-1}$ and $q_0=0.0$, 
implying an angular distance 
scale of $10.885 h_{50}^{-1}$ kpc arcsecond$^{-1}$ for 3C~280 and 
$9.700 h_{50}^{-1}$ kpc arcsecond$^{-1}$ for 3C~254. For reference, here 
are the angular distance scales and luminosity conversion 
factors for the currently most fashionable 
cosmology of $H_0=65$ km s$^{-1}$ Mpc$^{-1}$, $\Omega_M=0.3$, and 
$\Omega_{\Lambda} = 0.7$: the angular distance scale are 
$8.616 h_{50}^{-1}$ kpc arcsecond$^{-1}$ for 3C~280 and 
$7.838 h_{50}^{-1}$ kpc arcsecond$^{-1}$ for 3C~254. The 
luminosities quoted in this paper 
would decrease by a factor of the luminosity 
distance $D_L^2$: 0.63 for 3C~280 and 0.65 for 3C~254. 
 
\section{Individual Sources} 
 
3C~254 is a radio-loud FRII quasar with $z=0.734$ (Spinrad et al. 1985) 
at 11h 14m 38.5s +$40\degr$ 37$\arcmin$ 20$\arcsec$ (J2000). 
The radio source is very asymmetric, 
with its shortest hot spot-nuclear separation 
on the eastern side. Owen \& Puschell (1984)  
showed that the eastern lobe is extended roughly 
perpendicular to the source axis. An overdensity 
of galaxies near the source suggests the environment is a 
cluster or a group (Bremer 1997). This overdensity consists of  
seven objects within $10\arcsec$ ($97 h_{50}^{-1}$ kpc),  
with magnitudes within 3-5 magnitudes that of the 
central galaxy.  The magnitudes are consistent with 
passively 
evolving $L^*$ galaxies at $z=0.734$ with $z_{formation} > 2$ (Bremer 
1997).

3C~280 is an FRII radio galaxy with 
$z=0.996$ (Spinrad et al. 1985) at 12h 56m 57.1s $+47\degr$ 20$\arcmin$ 
20$\arcsec$ (J2000). 
It is one of the most powerful FRIIs known at radio frequencies. 
It has an extended ($11\arcsec$) emission line nebula ([OII]3727\AA), 
with a central peak and a loop  around the 
eastern radio lobe (McCarthy et al. 1987).  
 
Best's (2000) K-band images of both 
of these sources suggest a richness  
class 0 environment. At low redshift, richness class 0 
clusters of galaxies tend to have X-ray luminosities between $10^{43-44} 
\lum$. 
 
ROSAT observers reported extended emission from both of these 
sources (Hardcastle \& Worrall 1999; Crawford et al. 1999; Worrall et al. 
1994). 
For 3C~254, residuals with extent of $30-60\arcsec$ were estimated from 
PSPC images (Hardcastle \& Worrall 1999). Crawford et al. (1999) 
identifies excess emission out to at least $15\arcsec$ 
(145 kpc $h_{50}^{-1}$) in 3C~254, with 
12-19\% of the total X-ray luminosity in extended emission, 
corresponding 
to $L_x \sim 5-9 \times 10^{44} h_{50}^{-2} \lum$. For 3C~280, Worrall 
et al. (1994) reported PSPC evidence for a cluster extent of 
$18-65\arcsec$ ($200-700 h_{50}$ kpc). 
Hardcastle \& Worrall (1999) reported rest-frame 2-10 keV 
X-ray luminosities of $1.3 \times 10^{45} h_{50}^{-2} \lum$ and 
$2.8 \times 10^{44} h_{50}^{-2} \lum$ for 3C~254 and 3C~280 
respectively. 
Correspondingly, Hardcastle \& Worrall (2000) estimated rough ICM 
pressures 
of $\sim 7 \times 10^{-12}$ ($\sim 8 \times 10^{-13}$) erg cm$^{-3}$ for 
3C~254 (3C~280), 
assuming 7.7 (5.0) keV gas temperatures. In \S~\ref{rosat}, we 
have re-analyzed these ROSAT data for comparison purposes.

\section{Chandra Observations and Analysis}

Both 3C~280 and 3C~254 were observed with the Chandra 
ACIS-S in VFAINT datamode 
and TIMED readmode. We reprocessed the level 1 events file using 
updated gain maps from 
CALDB v2.9 (2001-10-22) and CIAO2.2\footnote{CIAO is the Chandra Interactive Analysis of Observations (CIAO). Version 2.2.1 was 
released on 13 December 2001, available at  
http://cxc.harvard.edu/ciao/.}. Only events with 
the ASCA grades of 0, 2, 3, 4, and 6 and a status$=0$ (``clean'') 
were used in the subsequent analyses.

3C~254 was observed for 29,668 kiloseconds (all corrections 
applied) on March 26, 2001. The observation was uncontaminated by 
flare activity, so the good time intervals, as supplied by the 
original pipeline, were used to extract the final clean data. 
 
3C~280 was observed for 63,528 (standard good time intervals) 
on August 27, 2001. A light curve of the background light between 
0.3-8.0 keV reveals several flares. Even the ``quiescent rate'' 
for the observations varied several times. For analysis 
of flare-free ``clean'' data, 
we only use the 18,944 seconds of data near the 
beginning of the observation, particularly for the analysis of 
extended sources. The signal to noise estimates of point sources in  
these clean data are extremely compromised relative to the total,  
so for analysis of the point sources, 
which are relatively unaffected by background (usually  
less than $\sim10\%$ of the total count rates), we also analyzed 
the full data set for 3C~280. For the point sources, the count rates 
and spectra from the full data set are statistically consistent with 
those extracted from the clean data, and therefore, for the spectral 
fits (only for the purposes of improving flux estimates of the 
point sources and for the approximation of their broadband spectra) 
were done with the full data set of 63,528 seconds. 
 
\begin{figure} 
\plotone{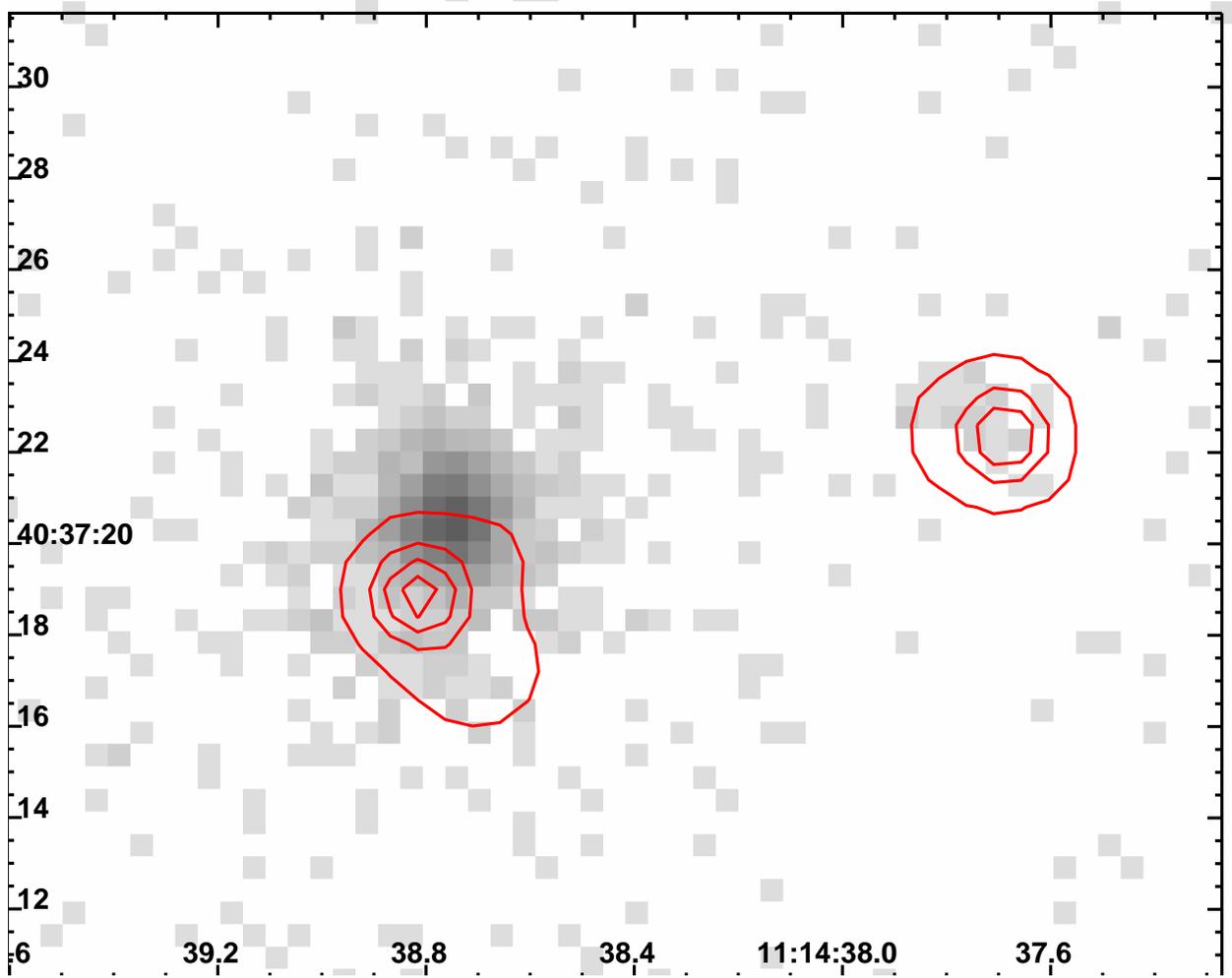} 
\caption{Radio contours overlaid on the X-ray greyscale images for 
3C~254 and 3C~280. 
The 3C~254 radio intensity 
map was made with the VLA by Liu, Pooley, \& Riley (1992), observed on 
March 21, 1989 with a resolution 
of $1.25\arcsec \times 1.25\arcsec$ at 4885.10 MHz. 
\label{figure:3c254}} 
\end{figure}

\begin{figure} 
\plotone{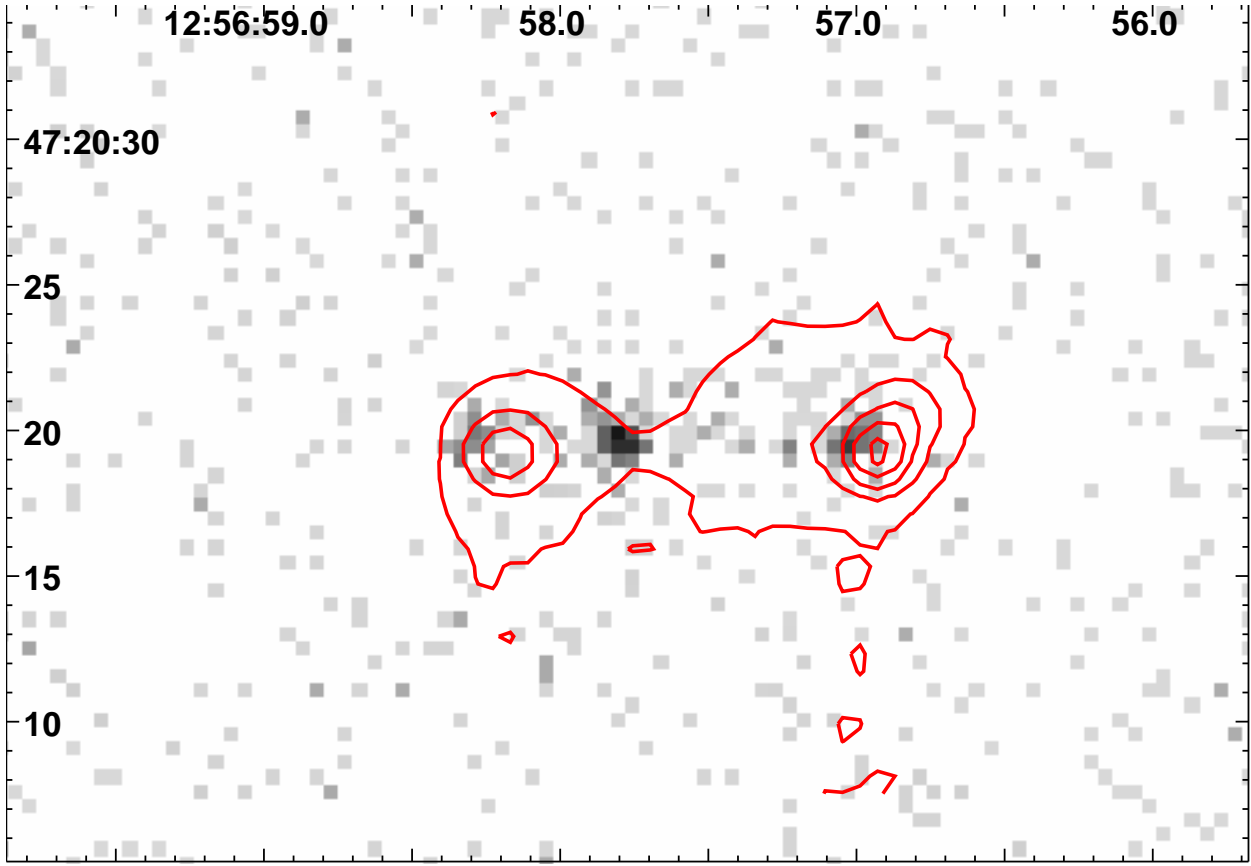} 
\caption{Radio contours overlaid on the inverse greyscale image of the 
Chandra X-ray image for 
3C~280. The 3C~280 radio map was obtained from Liu, Pooley, \& Riley 
(1992). The 
radio observation was made with the VLA 
on January 21, 1989, at 1464.90 MHz, with $1.20\arcsec \times 1.20\arcsec$ 
resolution. 
\label{figure:3c280}} 
\end{figure}

For 3C~254, we locate the position of  
a powerful central X-ray source, confirming the very 
short projected distance between the nuclear source and the western 
radio lobe seen by Owen \& Puschell (1984). The eastern radio hot spot was not  
detected (it may be too close to the central source),  
but the western hot spot has a faint X-ray counterpart 
(Figure~\ref{figure:3c254}). 
For 3C~280, we detect both the western and eastern hot spots as well as the 
nuclear source in the X-ray (Figure~\ref{figure:3c280}). There 
may be faint emission associated with 
the bridge; however, estimates of contamination from the scattering 
wings of the three point sources shows this emission to be very uncertain, 
whether evaluated in the restricted or the full dataset. 
Neither 3C~254 nor 
3C~280 (using only the cleaned dataset) showed any evidence for the putative 
cluster sources expected from 
the previous analyses of the ROSAT data for these same sources. Discussion 
of the Chandra limits are in \S~\ref{cluster_limits}. 
 
For the nuclei and the hot spot sources, 
we used a 4 pixel (1 pixel = 0.4920") 
radius aperture to extract counts. A 4 pixel radius 
encircles approximately 90\% of the energy at 1.7 keV, a correction 
which 
is not included in the fluxes and luminosities reported in 
Table~\ref{sources}. 
We also extracted 
a count estimate for the bridge region of 3C~280, excluding the 3 
point sources, for a region of $35 \times 10$ pixels ($17\farcs22 \times 
9\farcs42$), elongated exactly east-to-west, 
with 3 $r=4$ pixel regions excluded for a total 
area of of 150 square arcseconds. 
The background-subtracted count rates (0.3-10.0 keV)  
inside those apertures and the  
percentages of the total count rates in background counts are 
reported in Table~\ref{sources}. For each source, a PI spectrum 
was extracted and binned to at least 15 counts per PI bin. We used 
XSPEC v11.1 to fit a power law 
spectrum to the $0.3-8.0$ keV data, 
with Galactic absorption, 
fixed at $1.2 \times 10^{20}$ cm$^{-2}$ for 3C~280 and $1.8 \times 
10^{20} 
$ cm$^{-2}$ for 3C~254 (Dickey \& Lockman, 1990; Stark et al. 1992).  
The fit range is approximate, because of the width of the  
bins in each spectrum. A power law absorbed by the expected 
Galactic absorption adequately described the spectrum 
for all of these sources in terms of $\chi^2$. A thermal spectrum 
could also be fit to most of the spectra, except for those of the 
central sources of 3C~254 and 3C~280. The central source of 3C~280 seems 
to be unusually flat with a best-fit spectral index near 0.0; the 
cleaned data were consistent with that result, with a truncation near 
8 keV, but the poor statistics of that flare-affected data 
limit our ability to confirm the power law index from the full dataset -- the 
data from the clean set are consistent with the fit from the full, 
but cannot be used for an independent measurement of the slope of the 
power law.  
 
The central source of 3C~254 was affected by pile-up at 
about the $10\%$ level. 
Pile-up flattened the observed spectrum and 
produced a broad ``emission'' feature at around 2 keV. We first 
fit the spectrum without a pileup model to obtain a best first-guess 
estimate of the slope and normalization of $\alpha=1.3$ and 
normalization of $1.4 \times 10^{-4}$ respectively (the normalization units are 
photons keV$^{-1}$ cm$^{-2}$ s$^{-1}$ at 1 keV). The 
inclusion of the pile-up correction in XSPEC induces unstable fitting 
in local minima unless a reasonable first-guess is provided. 
With the XSPEC pileup model, with frame-time set to 
3.2 seconds, maximum number of piled-up photons set to 5, 
$\alpha_{pileup}=0.5$ and $g0=1$ 
(pileup model parameters fixed at recommended values),  
the best fit slope (photon index) was $\alpha=1.8\pm0.1$, 
$N_H=3.2 \times 
10^{20} (1.8-6.2)$ cm$^{-2}$, and normalization of $4.9 \times 10^{-4}$. 
The  
reduced $\chi^2=1.2$ for this fit with  
161 degrees of freedom obtains a null hypothesis 
probability 
of 0.05. This power law index somewhat exceeds the best-fit power-law 
index from ASCA data of $\alpha=1.67\pm0.04$ obtained by Sambruna, 
Eracleous, \& 
Mushotzky (1999), 
but the estimates overlap at the 90\% uncertainty level, which, given 
the uncertainties in the pile-up model, is not bad. 
 
For 3C~280 and 3C~254, thermal spectra from a 
gas with $kT \sim 2$ keV and metallicity of $\sim 30\%$ solar also 
adequately described the spectrum for the hot spot sources. The excess 
counts from the bridge of 3C~280 could be described by a cooler thermal 
spectrum of $1.3^{+0.2}_{-0.6}$ keV (with no constraints on the 
metallicity). As noted earlier, this source is likely to 
be heavily contaminated by scattered light from the hot spots and the 
nucleus so its flux is highly uncertain. Also, since  
it is not a point source, its spectral properties 
can be affected by how the background is handled.  
The inferred  broad-band fluxes and luminosities  
for all the sources depend somewhat on the assumption of 
a spectral model, whether power-law or thermal, so we 
report those properties under both model assumptions in  
Table~\ref{sources} for those sources with 
insufficient signal to distinguish between the two spectral models.

\section{Hubble Space Telescope Observations and Measurements 
\label{hst}} 
 
In order to detect or place limits on optical synchrotron emission from 
the radio hot spots, we retrieved HST Wide Field Planetary 2 (WFPC2)  
data from the HST Archive for both sources. We 
used co-added and cosmic-ray rejected 
data from both the CADC and the ECF HST data archives, and 
compared photometry from those co-added data with that from  
individual observations retrieved from the 
HST data archive. The co-addition process (with averaging) 
obtained consistent photometry  
for these observations. The gain for all observations 
was 7.0 electrons per DN (data number). The observations were reduced 
with standard calibration software using the most recent calibration 
files, switches, and software. No astrometric shifts were 
applied to the HST data. The IRAF/STSDAS task {\em invmetric} was 
used to derive the x and y positions corresponding to the right 
ascension and declination of the X-ray hot spots. 
 
The target 3C~254 was observed in a Planetary Camera (PC) 
snapshot program in both the F702W (R-band) and the F555W 
(V-band) filters. 
The nucleus was saturated, but the 
host galaxy was detected (Lehnert et al. 1999). We do not detect a 
counterpart 
for the radio hot spots in either image. We report a $3\sigma$ 
upper limit for a $1\arcsec \times 1\arcsec$  
aperture centered on the X-ray position 
of the western hot spot for the 
F702W observation in Table~\ref{table:hst}. 
 
\begin{figure} 
%\plotone{f3.eps}
\plotone{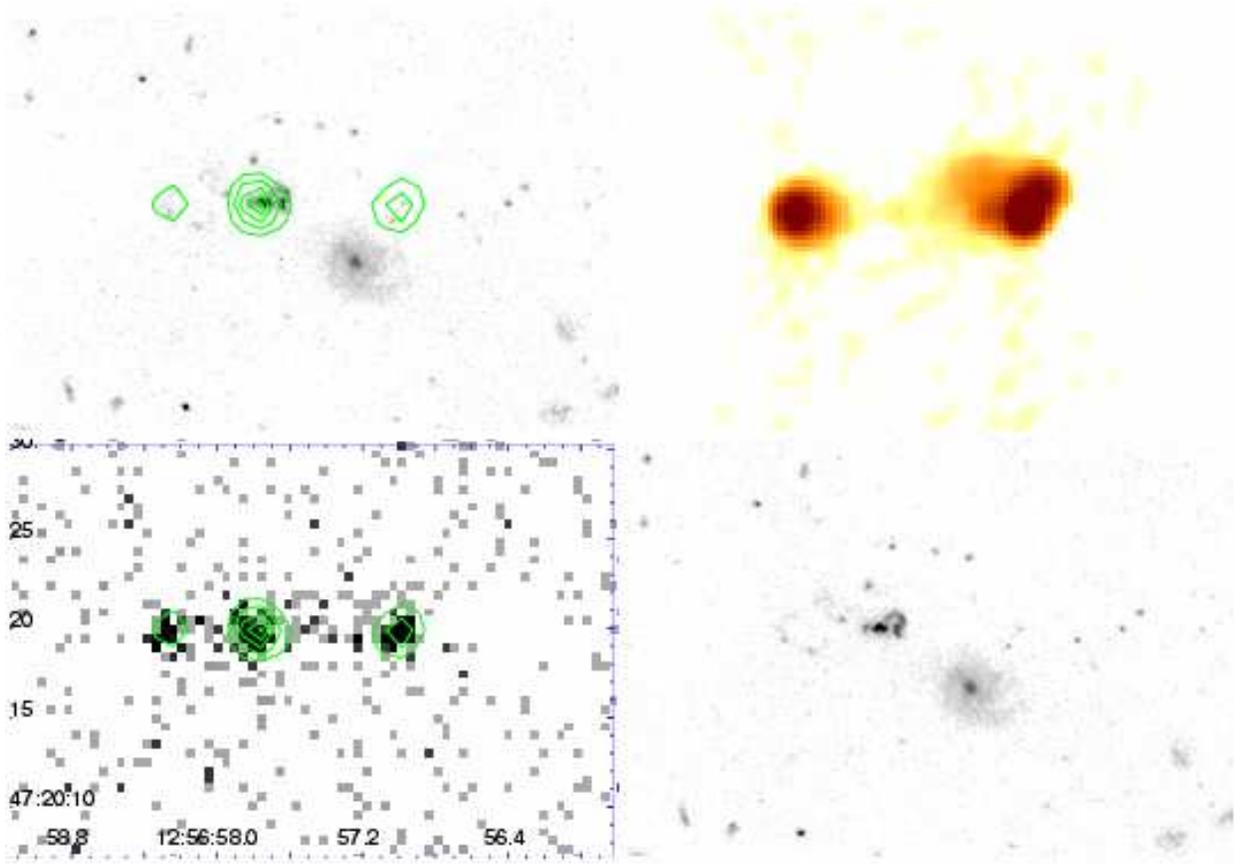} 
\caption{\label{figure:hst_radio_xray_3c280} 
This figure shows four observations of 3C~280 in the optical, 
radio, 
and X-ray, displayed with the same scale and orientation. 
The upper left figure is the HST WFPC2 image of 3C~280 with X-ray 
contours overlaid (sufficient to show the location of the core and the 
two hot spots); the lower right figure is the same grey-scale image with 
the X-ray contours omitted. The upper right image is the VLA radio image 
from 
Liu, Pooley, \& Riley (1992); the 
lower left image is the X-ray image with the same X-ray contours as the 
upper left HST image. We frame the lower left image with the coordinate 
system of all 4 images. The nucleus of 3C~280 seems to be coincident 
with 
feature ``a'' from Ridgway \& Stockton (1997).} 
\end{figure}

The target 3C~280 was observed in the Wide Field Planetary Camera 2 
(WFPC2), 
Wide Field 3, for a total of 8800 seconds in the F622W filter. Ridgway 
\& 
Stockton (1997) report the unusual appearance of the central galaxy. 
We find two point-source features 
within $0\farcs67$ and $0\farcs45$ of the western and eastern 
hot spot Chandra positions, and aligned with the peaks of the radio 
hot spots 
in the Liu, Pooley, \& Riley (1992) maps 
(Figure~\ref{figure:hst_radio_xray_3c280}). 
We extracted background-subtracted 
aperture ($r=0\farcs3$)  
photometry at the positions of the HST point sources. Using 
the basic flux calibration information from  
the header (the PHOTFLAM keyword) we 
converted 
the observed count rates (DN sec$^{-1}$)  
to flux densities in nJy at 6200 \AA. We 
corrected 
these fluxes for geometric distortion ($\sim1\%$) (Casertano 
\& Wiggs 2001), charge transfer 
inefficiency 
($\sim2-3\%$), and aperture ($\sim16\%$) (Whitmore \& Heyer 2002; 
HST Data Handbook.) 
 
The optical flux densities of these point sources are consistent with 
optical synchrotron emission from the radio and X-ray hot spots, but 
are very unlikely to be produced by  
synchrotron self-Compton (SSC) or inverse Compton (IC) of the 
cosmic microwave background (CMB).  We will 
discuss this topic further in \S\ref{hot spots}.

\section{Evaluation of Extended X-ray Emission} 
 
The distribution of X-ray emission in clusters of galaxies  
is typically extended, with a core radius of $\sim100-200h^{-1}$ kpc,  
and is centered near the center of mass of the cluster. The X-rays 
arise from an optically thin plasma confined by the gravitational 
potential of the cluster; the temperature of the gas is indicative 
of the depth of the potential. Cluster luminosities are typically 
$10^{43}-10^{45}$ erg s$^{-1}$.  
Cluster temperatures run between 
2-15 keV and typical central densities range between $10^{-2}$ and 
$10^{-4}$ cm$^{-3}$.  
In the following section, we describe the analysis performed 
in order to place constraints on the presence of an extended 
component arising from a cluster of galaxies  
centered on these radio sources. 
 
\subsection{Radial Profile of 3C~254} 
 
A visual inspection of the 3C~254 X-ray image reveals a faint halo around the 
central bright source. Here we show this halo is consistent with 
being part of the point spread function (PSF) of a very bright 
Chandra X-ray source. 
 
\begin{figure} 
%\plotone{psf_comp.ps}
\plotone{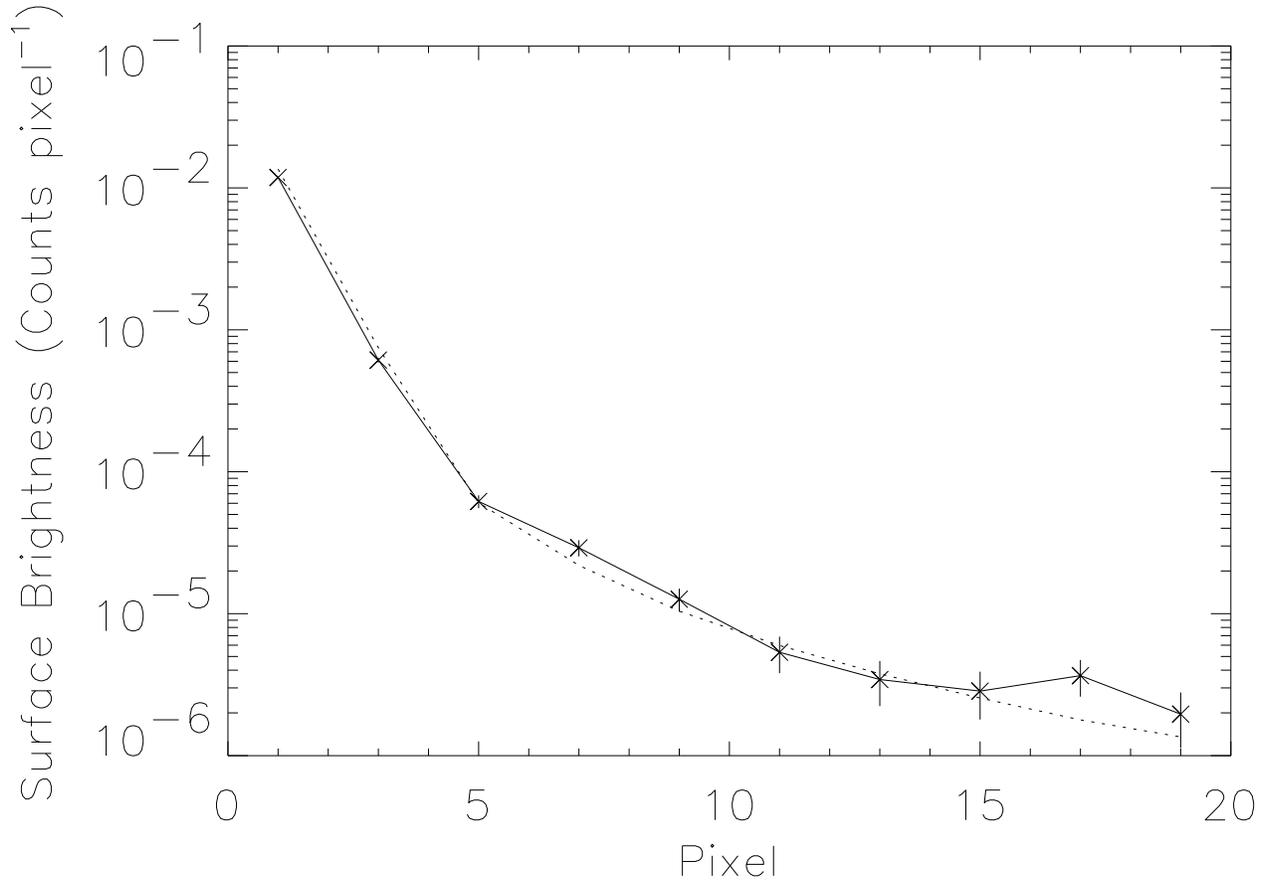} 
\caption{\label{figure:PSF} The radial profile of the 
relative surface brightness, in counts per  
detector pixel ($\sim0.5\arcsec \times 0.5\arcsec$) per  
second, normalized,  
of the central source of 3C~254 (solid line) and a radial  
profile of the weighted  
distribution from summed PSFs from the Chandra PSF library (dotted 
line). No 
significant excess is seen in the central $10\arcsec$ (20 pixels).} 
\end{figure}

We constructed a model PSF from the library of Chandra PSFs by 
extracting PSFs for the on-axis position on the ACIS-S from the library for 
six different energies: 0.80, 1.25, 2.25, 3.50, 4.50, and 5.50 
keV respectively. Based on the energies of photons in the central 
X-ray source of 3C~254, each PSF was weighted 
to construct a summed, normalized PSF. 
In Figure~\ref{figure:PSF}, the resulting radial profile of the 
predicted PSF is 
plotted against the radial profile of the  
central source, extracted identically. 
Inspection of this plot reveals that there is no detectable excess 
within $10\arcsec$ of the central source over the halo of a bright 
point source. We therefore exclude a region of radius $12\arcsec$ 
around the X-ray bright nuclear source of 3C~254 when placing  
a limit on cluster emission around this radio quasar (\S5.2).

\subsection{Upper Limits to Cluster Emission and to Density Profiles 
of Hot Gas \label{cluster_limits}} 
 
\begin{figure} 
%\plotone{regions.ps}
\plotone{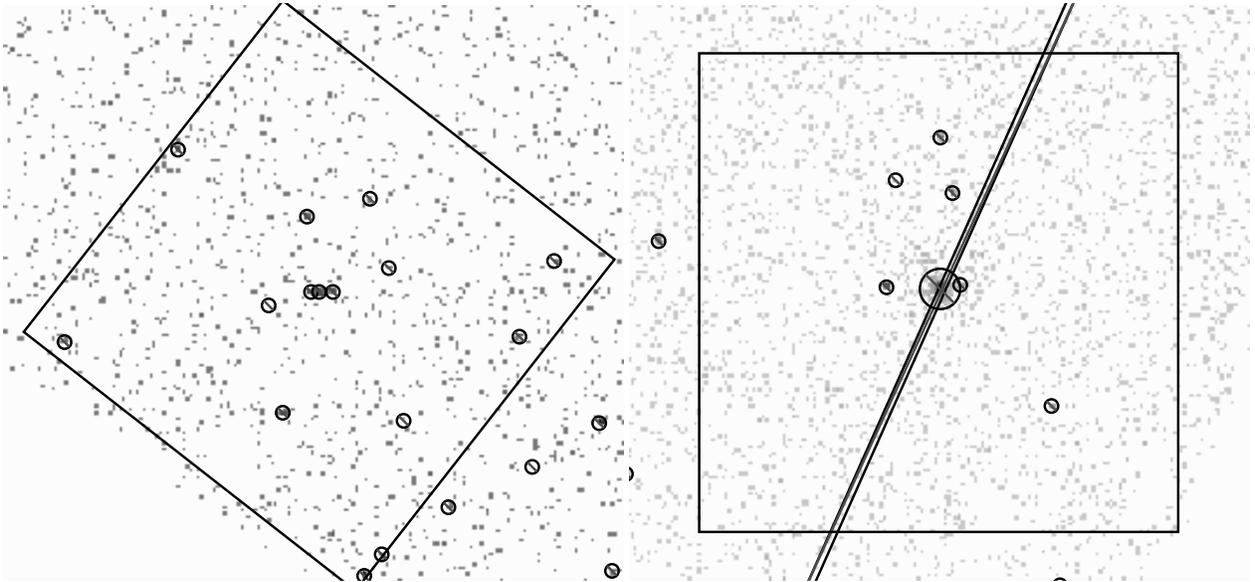} 
\caption{Image of a portion of the Chandra ACIS-S3 chip (binned by 4) 
 Êfor 3C~280 (left) and 3C~254 (right). ÊIn each image, the large box 
 Êshows the region used in surface brightness fitting. ÊThe circles 
 Êwith slashes show the excluded regions (usually sources found by 
 Ê\em{wavdetect}. ÊFor 3C~254, the diagonal box shows the region 
 Êexcluded due to the readout streak. 
 Ê\label{figure:regions}} 
\end{figure}

We used SHERPA, CIAO's fitting engine,  
to fit the extended, flare-filtered data within $\sim 3 
h_{50}^{-1}$ Mpc for both targets between $0.3-2.0$ keV with the point 
sources detected by the CIAO routine {\em wavdetect} masked out (see 
Figure~\ref{figure:regions}). ÊFor 3C~254, we also masked out a larger 
region ($\sim 12 \arcsec$) around 3C~254 and a region around the 
readout streak. ÊThe $0.3-2.0$ keV energy range maximized our 
sensitivity to soft thermal emission while minimizing the contribution 
from the particle backgrounds. Images were binned by a factor of 4 
($\sim2\arcsec$ pixels) to increase the S/N of each pixel and to reduce the 
computation time for fitting. Ê 
Using the Cash statistic to assess the relative 
goodness of fit, we fit a circular beta-model profile to the 
observed surface brightness $S_x$ with the 
function $S_x = S_0 (1 + (\theta/\theta_C)^2)^{-3\beta + 1/2}$ and a 
constant background. $S_0$ is the normalization, $\theta$ is the 
angular separation from a center (assumed to be aligned with the 
central X-ray point source), $\theta_C$ is the core radius, and 
$\beta$ is the slope of the traditional cluster beta function  
(typically $\sim2/3$.) The beta-model slope $\beta$ and the core 
radii ($\theta_C$) we tested span a range of standard cluster and group 
values. We then derived the $3\sigma$ upper limit to the best fit 
central surface brightness using the SHERPA {\em projection} command 
(Table~\ref{table:3c254_results} and \ref{table:3c280_results}) for 
each pair of $\beta-\theta_C$ values. ÊTo infer a limit on 
the cluster luminosity and central density, we derive the limit on the 
total count rate within an aperture of radius  
$2 h_{50}^{-1}$ Mpc, then convert that to a 
bolometric luminosity limit by assuming a temperature 
within 1.0 keV of the Markevitch (1998) $L_x-T_x$ relation. 
Approximate limits on $L_{bol}$ are $2.5 \times 10^{43} h_{50}^{-2}$ 
erg s$^{-1}$, for $N_H=1.2 \times 10^{20} ~\rm{cm}^{-2}$ and $1.87 
\times 10^{20} ~\rm{cm}^{-2}$ (Dickey \& Lockman, 1990; Stark et al. 1992) 
for 3C~280 and 3C~254, respectively 
(Table~\ref{table:densities}).

We convert the central surface brightness  
to a central electron number density, following the 
derivations for the emission measure from Equation 5.68  
in Sarazin (1988) where $n_e/n_p = 1.209$ for a fully ionized plasma 
of solar abundances. For a spherically symmetric plasma, the  
central electron density   
in units cm$^{-3}$ is: 
\begin{equation} 
n_{e} = \sqrt{\frac{(n_e/n_p) \, 4 \pi^{1/2} \, k \, (1+z)^2 \, S_0 \Gamma(3\beta)} 
                  {3.0856 \times 10^7 \, r_c \, \Gamma(3\beta-1/2)}} 
\label{eq:density} 
\end{equation} 
where  $r_c$ 
(core radius) is in kiloparsecs.  
We define $k$ here to be XSPEC normalization of a Raymond Smith (or 
Mekal) thermal spectrum divided by the Chandra count rate in the source 
spectrum. In this definition, $\int n_e n_H dV = k \times CR 
\times 10^{-14} / ( 4 \pi (D_A (1+z))^2) $ where $CR$ is 
the total count rate (counts sec$^{-1}$) from the volume $dV$, $D_A$ is 
the angular size distance to the source in $cm$, $n_e$ is the electron 
density ($cm^{-3}$),  
$n_H$ is the hydrogen density (cm$^{-3}$)  and $V$ is volume in 
units $cm^3$ (XSPEC 11.1, Raymond and Smith model.) The  
normalization per unit count rate $k$ depends 
on bandpass, temperature, metallicity, and absorption column. The quantity $S_0$ is 
counts sec$^{-1}$ sr$^{-1}$.  
We compute the total {\em projected} emission 
measure along our line of sight to the cluster  
inside an aperture of $2 h_{50}^{-1}$ Mpc, 
derived assuming Galactic absorption and 
Raymond-Smith models in XSPEC with the response matrices appropriate 
to the center of the ACIS-S chip.  
 
The normalization $k$ for the 0.3-2.0 keV 
count rates from a 1.0 keV 
plasma with a 30\% solar metal abundance (typical of a 
cluster of galaxies at $z<0.8$ (Donahue et al. 1999)) 
are $1.43 \times 10^{-2}$ in units of XSPEC  
normalization per Chandra count rate  
(counts sec$^{-1}$) for 3C~280 and $1.08 \times 10^{-2}$  for 3C~254. 
The conversion factor from central surface brightness  
(in units of counts sec$^{-1}$ arcmin$^{-2}$) to bolometric 
luminosity for the same plasma, $\beta=0.7$ and $r_c=250 h_{50}^{-1}$  
kpc, are  
    $5.02 \times 10^{46}  h_{50}^{-2} \lum/ (\rm{count \, sec^{-1} \, arcmin}^{-2}$)  
and $2.28 \times 10^{46}  h_{50}^{-2} \lum/ (\rm{count \, sec^{-1} \, arcmin}^{-2}$) for 
3C~280 and 3C~254 respectively.  
 
The $3\sigma$ upper limits on the central electron density,  
given $\beta$, $r_{core}$, and $k(T_x, Z_\odot, N_H)$, are  
derived using Eq.~\ref{eq:density}.  
In Table~\ref{table:densities}, we report  
$3\sigma$ 
limits on the bolometric X-ray luminosity ($L_{bol}$) inside $2000 
h_{50}^{-1}$ kpc, 
the central electron density ($n_e$), and the electron density ($n_e(r)$)  
at the hot spot 
radii, $r$,  
given $r_{core}=250 h_{50}^{-1}$ kpc and $\beta \sim 0.7$ 
(typical of clusters of galaxies).  We estimate $n_e(r)$ to compare 
to the electron density $n_w(r)$ computed by WDW97.  
We also explored the same limits for a range of $\beta = 0.55-0.75$ and 
core radius $r_{core} = 150 - 350 h_{50}^{-1}$ kpc. All combinations 
of those two parameters predicted central electron densities of 
$\sim(2-14) \times 10^{-4} h_{50}^{1/2}$  and not much smaller densities 
at $78 h_{50}^{-1}$ kpc (since $r<r_{core}$), where the Wellman et al. 
(1997a) predictions were made based on the radio sources.

WDW97 predict two sets of ambient gas densities; 
one set assuming minimum energy conditions to obtain the magnetic 
field strength in the radio bridge region, and one set assuming that 
the magnetic field strength in the radio bridge is about 1/4 of 
the minimum energy value, as suggested by comparisons of the 
radio and X-ray determined ambient gas densities around 3C 295 
(Perley \& Taylor 1991) and 
Cygnus A (Carilli et al. 1991).  This magnetic field 
strength is parameterized by $b=B/B_{min}$. 
WDW97 assumed a value of 
Hubble's constant of $100 \hbox{ km s}^{-1} \hbox{ Mpc}^{-1}$. 
For the case $b=1$, synchrotron cooling dominates over inverse 
Compton cooling with the cosmic microwave background radiation, and 
the radio determined ambient gas density scales as $H_o^{12/7}$ 
(see the footnote to Table 4 of Guerra, Daly, \& Wan 2000), 
so for value of $H_o$ adopted here, all of the ambient gas densities 
listed in Table 1, column 9 of WDW97 
decrease by a factor of 0.3.  Thus, the 
ambient gas density a distance of about $78 h_{50}^{-1}$ kpc from the nucleus 
of 3C~254 
is predicted to be 
about $1.5 \times 10^{-3} \hbox{ cm}^{-3}$.  The ambient gas 
density for the western lobe of 3C~280 at a distance of about 
$76 h_{50}^{-1}$ kpc from the center of the radio source is 
predicted to be 
about 
$7 \times 10^{-3} \hbox{ cm}^{-3}$, while that for the 
eastern lobe at a distance of about 62 kpc from the radio source 
center is about $2.4 \times 10^{-3} \hbox{ cm}^{-3}$ for $b=1$. 
These ambient gas densities are significantly larger than the 
$3\sigma$ upper bounds provided by the Chandra data for 
both sources. For $n_W$ for 3C~254, $b=1$ estimates of $n_W$  
could be consistent with the X-ray upper limits if  
$r_c < 150 h_{50}^{-1}$ and $\beta > 0.7$, values that are  
more consistent with belonging to groups than clusters. 
 
For the case $b = 1/4$, the radio predicted ambient gas 
densities are consistent with the X-ray bounds for 
most values of $\beta$ and $r_c$.  We report the converted densities 
($n_W$) in Table~\ref{table:densities} For $b=1/4$ 
the bridge magnetic energy density for the bridges of 3C~254 and 
3C~280 are comparable to the energy density 
of the cosmic microwave background, so inverse Compton cooling 
is as important in these regions as synchrotron cooling, and the 
radio-predicted ambient gas density scales as $H_0^{20/7}$ 
(see the footnote to Table 4 of Guerra, Daly, \& Wan 2000). 
For the case $b=1/4$, the predicted ambient gas densities are 
obtained from column 11 of Table 1 of WDW97 
and multiplied by the factor 0.14 $=(0.5)^{20/7}$ to account for 
the value of $H_0$ adopted here.  The radio-predicted ambient 
gas densities are then about $2.1 \times 10^{-4} \hbox{ cm}^{-3}$ for 
the western lobe of 3C~254, and about $8.3 \times 10^{-4} 
\hbox{ cm}^{-3} h_{50}^{20/7}$ for both the 
western and eastern lobes of 3C~280.    These gas 
densities predicted by 
Wellman et al (1997a) ($n_W$) for $b=1/4$ 
are listed in Table~\ref{table:densities}. For both sources, 
the radio-predicted ambient gas density is 
consistent with the upper bound provided by Chandra for $b=1/4$. 
 
The radio predicted ambient gas density has a very strong 
dependence on $H_0$ and thus also the coordinate (or proper) 
distance ($a_0r$) to the source.  When inverse Compton 
cooling with CMB photons is important, the 
ratio of 
the radio determined ambient gas density $n_W$ to the  
X-ray determined value $n_e$ is  
$n_W/n_e \propto (a_0r)^{-2.36} \propto  
h_{50}^{2.36}$.  If we shift from the cosmology assumed 
here, $H_0 = 50$ km s$^{-1}$ Mpc$^{-1}$ and $q_0 = 0$, to  
a cosmology with $H_0=65$ km s$^{-1}$ Mpc$^{-1}$,  
$\Omega_m = 0.3$, and $\Omega_{\Lambda} = 0.7$, the ratio 
between the predicted value by Wellman et al (1997a) and the 
$3\sigma$ upper limit from the Chandra observations is increased 
by a factor of about $(0.8)^{-2.36} \simeq 1.7$.   
This effect does not 
change our conclusion presented above for 3C~254, but it does affect the   
comparison between the radio-predicted ambient density 
and the upper limit on the density from the X-ray observations 
of 3C~280, where the radio-predicted ambient density may be 
in conflict with the X-ray observations at about the $1\sigma$ 
confidence level.  The radio predicted ambient gas density 
is increased by the ratio of the coordinate distances, (0.8),  
to the power 
2.86, leading to a predicted ambient gas density for 3C 280  
(see Table 5)  
of about 
$(16 \pm 5.5) \times 10^{-4}$ cm${}^{-3}$, whereas the upper bound placed by 
the Chandra observations increases by a factor of the square root of the 
ratio of 
the coordinate distances, $(0.8)^{-0.5}$  
to about $10 \times 10^{-4}$ cm$^{-3}$.    
 
An additional test could be made using our limits on the 
temperature of any putative X-ray-emitting gas distribution if the 
ambient gas pressure based on 
the radio properties of 3C~254 and 3C~280 could have been determined, 
as it has for many other 
high-redshift 3C sources (see Table 1 of Wan et al. 2000). 
The luminosity limits are extremely 
stringent for both sources, corresponding to bolometric luminosities of 
$\sim 3.5 \times 10^{43} h_{50}^{-2}$ and corresponding 
temperatures $T_x \sim 1.5$ keV. 
 
These observations rule out the presence of a hot, massive cluster  
with a bright intracluster medium for these particular radio sources. 
Specifically, we definitively rule out the presence of a cluster with the 
luminosity and distribution of X-ray plasma like the one 
containing Cygnus A and like that of the putative clusters  
associated with these sources, as inferred from 
ROSAT data (see more in \S~\ref{rosat}). 
 
Our conclusions are consistent with the recent findings from an 
optical study by  
Harvanek \& Stocke (2002) that found that FRII-selected clusters 
are less rich (Abell richness class 0-1) than their X-ray selected 
cousins, which tend to contain FRI sources exclusively. In their 
study, based on a review of radio structures and surrounding 
galactic density for 3CR radio galaxies between $z=0.15-0.65$, 
they conclude 
that radio sources with a large bending angle cannot be used as 
signposts for rich clusters around powerful radio sources.

\subsection{ROSAT PSPC and HRI Fluxes \label{rosat}} 
 
In this section, we compare the Chandra observations with both 
earlier results from the literature and with the archived ROSAT 
observations, in order to try to understand why clusters were 
reported around these objects. We find that the total fluxes measured 
by ROSAT and Chandra within 
$120\arcsec$ are in agreement for 3C~280, and that the nuclear source 
for 3C~254 has likely varied in the last five years. Therefore, at least 
for 3C~280, the extended emission inferred from the ROSAT data is 
explained by faint point sources.  
 
3C~280 was observed by the ROSAT PSPC in 1997 for 47.6 kiloseconds. 
Worrall et al. (1994) found that the PSPC radial profile was best fit 
by a point source plus a $\beta$-model, although they could not  
constrain the core radius very well.   
They stated that 60\% of the net counts in 
the 0.2--1.9 keV band were from an extended component around 3C~280 and 
a flux density of 1.7 nJy at 1 keV from the extended component. 
Although not actually given, their total flux density can be computed 
using quantities in Table~4 of Hardcastle \& Worrall (1999).  The 
total flux density at 1 keV is 6.2 ($\pm$ 3.6) nJy within a source 
radius of 120$\arcsec$ (uncertainties were high for the extended 
component).  They also noted a possible asymmetry in the X-ray 
emission.  
 
\begin{figure} 
\plotone{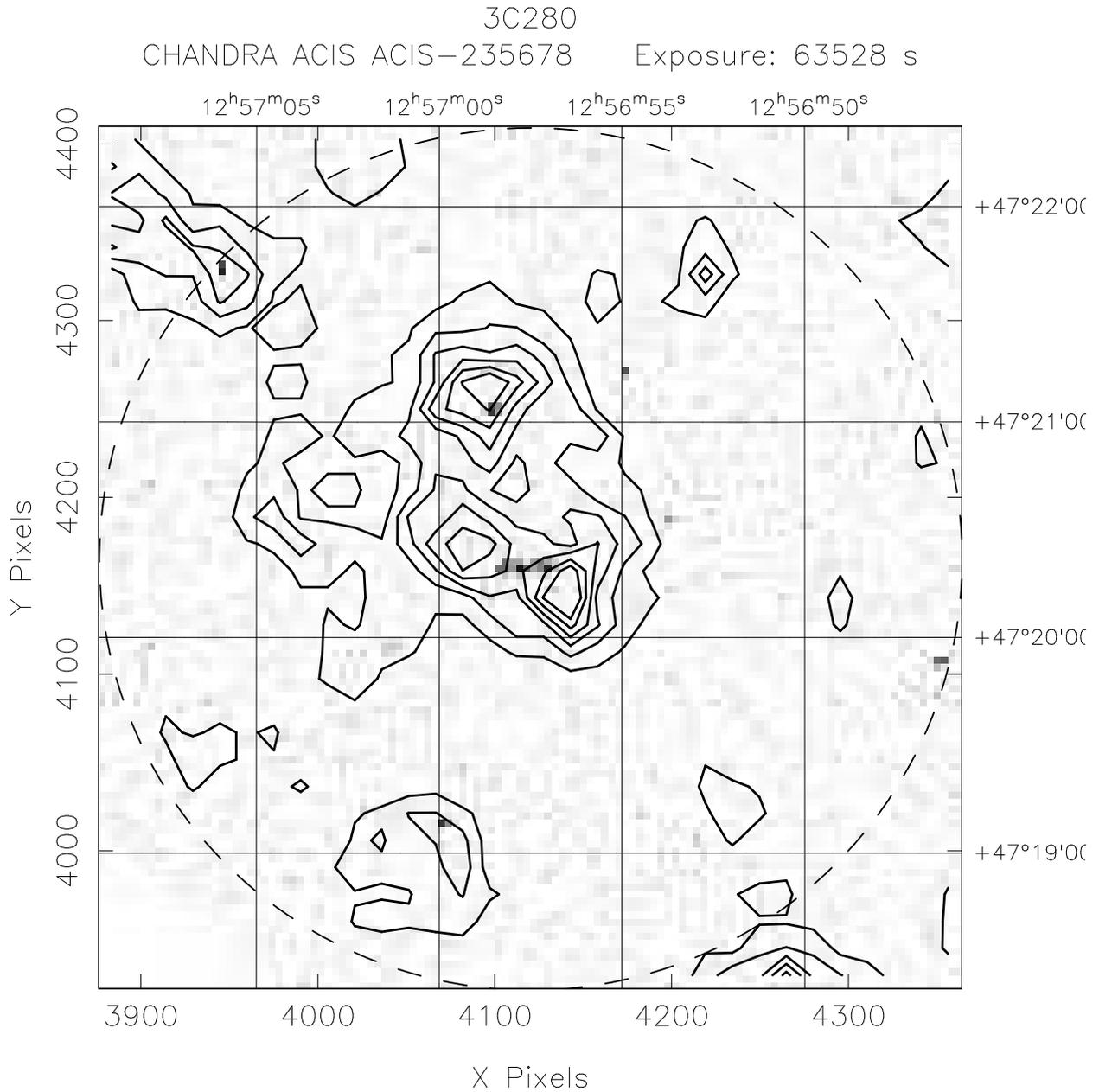} 
\caption{ 
  Chandra ACIS-S3 image (binned by 4) overlaid with ROSAT PSPC 
  0.5--2.0 keV X-ray contours.  The contours begin at 1.5 times the 
  background level and increase in steps of 0.5 times the background 
  level thereafter.  The dashed circle indicates the 120$\arcsec$ 
  source radius used by Worrall et al. (1994). 
  \label{figure:3C280_overlay.ps}} 
\end{figure}

Using the same PSPC data, we cleaned three episodes of high background 
from the data for net 42.7 kiloseconds (corrected for dead time).  We 
adaptively smoothed the 3C~280 events with energies between 0.5--2.0 
keV (the lower limits means we cut out much of the soft X-ray 
background) to a minimum of 10 counts in the smoothing kernel. 
Figure~\ref{figure:3C280_overlay.ps} shows the resulting  
PSPC contours overlaid on the $4 \times 4$ binned  
Chandra 63 ks image. 
 
The emission has at least three peaks within the $120\arcsec$ source 
radius used by Worrall et al. (1994).  The northern peak is definitely 
seen in the Chandra image as a separate point source.  Other, fainter 
sources in the Chandra image may also be contributing to the total 
PSPC flux.  Worrall et al. (1994) considered all these emission 
regions as one source.   
 
As a check, we compare the total PSPC flux within a $120\arcsec$ 
region around 3C~280 to the summed flux of Chandra point sources 
within the same region.  First, we extract the PSPC spectrum for an 
aperture of $120\arcsec$ around 3C~280 with a background from an 
annulus about 3$\arcmin$-4.5$\arcmin$, masking out several sources in 
the background.  A custom ARF and standard PSPC RMF were used; 
the spectrum was binned to 8 energy bins.  The resulting spectrum was 
fit to an $\alpha=2$ power law model in XSPEC 11.2 leaving only the model 
normalization as a free parameter.  $N_H$ was fixed at the Galactic 
value (Dickey \& Lockman 1990; Stark et al. 1992). 
We derive a 0.5-2.0 keV flux of $2.4 \pm 1.5\times 10^{-14}$ 
$\flux$, which corresponds to a flux density of $6.6 \pm 4.1$ nJy at 1 
keV.  This value is statistically the same as the 6.2 nJy 1~keV flux  
found by Worrall et al. (1994).  
A similar analysis of the ROSAT HRI data (observed for 53 ks in 1997) 
gives an HRI count rate of $1.1 \times 10^{-3}$ ct s$^{-1}$ which 
corresponds to a flux of $2.2 \times 10^{-14}$ $\flux$ or 6.1 nJy at 1 
keV.  
 
The sum of the estimated flux densities from Chandra sources within 
$120\arcsec$ of 3C280,  
including five background sources,  
is $\sim 4.9 \pm 0.6$ nJy (with an 
approximate $10\%$ correction for the flux outside the 4 pixel apertures). 
The sum of the point sources and bridge source within $120\arcsec$ of the 
3C~280 nucleus is statistically consistent with but somewhat 
lower than the ROSAT PSPC and HRI estimates from the original data. 
So most, if not all the ROSAT flux can be accounted for by the point 
sources in the Chandra observation. 
 
3C~254 was observed by the ROSAT PSPC for 16 ks in 1997 and with the 
ROSAT HRI for 29 ks in 1996.  Hardcastle \& Worrall (1999) report a 
total (extended and point-like) 1 keV flux density of $\sim250 \pm 13$ 
nJy.  They found that the radial profiles of both the PSPC and HRI 
data showed evidence for extended emission, although the HRI data gave 
a somewhat smaller and less luminous extended component. 
 
We analyzed the PSPC observation of 3C~254 similarly to 3C~280 and, 
assuming a power law index of 1.0,  
found a flux density at 1 keV of $168 \pm 10$ nJy, somewhat smaller 
than the Hardcastle \& Worrall estimate.  Our HRI flux density is also 
somewhat lower at 201 nJy.   
However, the Chandra estimate of the 1 keV 
flux density is $\sim300$ nJy, larger than either of these values. The 
small difference in assumed power law indices assumed (Hardcastle \& 
Worrall 1999 assume a power law index of 0.8) does not account for 
the variation.    
The comparison between the ROSAT and Chandra fluxes shows   
what is likely to be long-term X-ray variability of the radio 
quasar.

Table~\ref{table:rosat} summarizes our ROSAT results.

\section{X-ray Emission Associated with the Radio Hot Spots 
\label{hot spots}}

The X-ray emission that is coincident with the radio hot 
spots could be produced by the synchrotron self-Compton process 
(SSC), by synchrotron radiation, by 
inverse Compton scattering with external photon fields,  
by a combination of these processes, or by some other process. 
SSC and synchrotron radiation are discussed in 
\S 6.1 and 6.2.  Considerations of inverse Compton scattering of an 
external photon field, such as the cosmic microwave background 
radiation (CMB), allow a lower limit to be placed 
on the magnetic field strength 
of the region, which is discussed in \S 6.3.  Constraints on the 
bridge magnetic field strength of 3C~280 are also discussed in this 
section.

\subsection{The Synchrotron Self-Compton Process} 
 
Sources with Chandra detections of hot spot emission that 
is most likely due to SSC include 3C 295 (Harris et al. 2000), 
Cygnus A (Wilson, Young, \& Shopbell 2000), 
3C 123 (Hardcastle, Birkinshaw, \& Worrall 2001) 
and 3C 207 (Brunetti et al. 2002).   
 
SSC emission is most important in bright small regions, 
such as radio hot spots, 
while IC upscattering of external photons, such as those that 
comprise the CMB, is important in the more diffuse lobe and 
bridge regions of the source.  The dominant mechanism in any area 
is determined by the ratio of the local photon energy densities. 
IC scattering of local synchrotron photons, known as 
synchrotron self-Compton emission (SSC), will dominate 
over scattering of ambient 
photons, such as those of the CMB, when 
the local energy density of synchrotron photons is larger than 
the energy density of other photon fields.  For a uniform 
spherical synchrotron emitting region 
with bolometric luminosity L, radius r, and volume V, the local 
photon energy density is approximately 
$u_{s} \approx 0.75  Lr/(cV)$ (Band \& Grindlay 1985). 
The bolometric luminosity $L= \int L_{\nu} 
d\nu$ 
can be written in terms of a bolometric correction factor $\kappa_{\nu}$, 
where $\kappa_{\nu} =L/(L_{\nu}\nu)= f/(f_{\nu_o}\nu_o)$;  
$f$ is the total flux, $f=\int f_{\nu_o} \nu_o$,  
and $f_{\nu_o}$ is the flux density 
detected at frequency $\nu_o$.  The energy density 
of synchrotron photons in a region with diameter $2r$ and corresponding 
angular diameter $\theta$ is  
$u_s \approx 
(9 \kappa_{\nu} f_{\nu_0} \nu_0 (1+z)^4))/ (c \theta^2)$.   
Writing the  
flux density $f_{\nu_0}$ in units of Jy, $\theta$ in  
units of arcseconds, and the frequency $\nu_0$  
in units of GHz, the energy density of synchrotron photons 
of a radio hot spot is  
\begin{equation} 
u_s \approx 1.3 \times 10^{-13} \kappa_{\nu} (1+z)^4  
f_{\nu_0} \nu_0 \theta^{-2} 
\, {\rm erg \, cm}^{-3}. 
\end{equation}  
 
A minimum local photon energy 
density due to external sources is provided by the microwave background 
radiation, which has an energy density of $u_{cmb} = 
4.19 \times 10^{-13} (1+z)^4 
~\hbox{erg cm}^{-3}$ assuming a zero redshift 
temperature for the black body radiation 
of 2.728 K (Fixen et al. 1996). 
The ratio of the energy densities at the location of 
the radio hot spot is 
\begin{equation} 
\left({u_s} \over {u_{cmb}}\right) 
\approx 0.3~ \kappa_{\nu} ~\left({f_{\nu_o}} \over {Jy} \right) 
\left({\nu_o} \over {GHz}\right)~\left({\theta} \over 
{arcsec}\right)^{-2}, 
\end{equation}  
which is independent of choices of global 
cosmological parameters, redshift, and minimum energy assumptions. 
Thus, only in small, bright, radio-emitting regions will SSC  
dominate over IC with CMB photons and other ambient photon fields. 
 
Published radio observations do not resolve the hot spots of 
either 3C~254 or 3C~280.  The highest resolution image of 
3C~254 available is a 4.885 GHz image obtained at $0.5\arcsec$ 
resolution (Owen \& Puschell 1984).  The western hot spot 
has a flux density of about 0.13 Jy, and an angular size 
less than about $0.5\arcsec/\sqrt{ln2}$.  Assuming that the angular diameter of 
the hot spot is less than $0.5\arcsec/\sqrt{ln2}$, equation (2)  
indicates $u_s/u_{cmb} \gtrsim  1.5$ 
assuming a value of $\kappa_{\nu} \sim 3$, 
typical of sources with confirmed SSC phenomena and 
observed at similar frequencies, such as 
3C~295 (Harris et al. 2000) and Cygnus A (Wilson, Young, \& 
Shopbell 2000).  This energy density ratio indicates that 
SSC is likely to be more important than  
IC with external photon fields within the hot spot  
region. We stress that 
equation (2) is only a rough approximation obtained  
assuming that the hot spot is a uniformly emitting sphere 
whose angular size and radio spectrum can be accurately determined. 
On a slightly larger scale, about $1.25\arcsec$ in diameter, the 1.4 GHz 
flux density of the western hot spot of 3C 254 is about 0.7 Jy 
(Liu, Pooley, \& Riley 1992).  Substituting a value of  
$\theta$ of $1.25\arcsec/\sqrt{ln2}$ into equation (2) implies 
that $u_s/u_{cmb} 
\sim 0.4$ suggesting that on scales smaller than $1\arcsec$ or 
so SSC will be more important than  
inverse Compton scattering with CMB photons, whereas on scales 
larger than an arcsecond, inverse Compton scattering with  
CMB photons will be more important in producing X-rays than SSC.   
 
For 3C~280, the highest resolution 
radio data available is the 1.4 GHz map of Liu, Pooley, 
\& Riley (1992); the angular resolution is about  
$1 \arcsec$.  Assuming that the angular diameter 
of the source is 
less than about $1\arcsec/\sqrt{ln2}$,  
and adopting $\kappa_{\nu} \approx 3$, 
equation (2) indicates that 
$u_s/u_{cmb} \gtrsim 1.5$ and 0.7 for the west and 
east hot spots 
of 3C~280, which have 1.4 GHz radio flux densities of 
about 2 and 0.9 Jy, respectively. 
Thus, SSC is likely to dominate over inverse Compton 
scattering within the central $1\arcsec$ diameter of the hot 
spot, while inverse Compton scattering with ambient photon fields 
is likely to dominate outside this region.  Both processes may 
contribute to the X-ray emission, and SSC  
could be an important process 
in the hot spot regions of 3C~254 and 3C~280. 
 
A rough estimate of 
the SSC X-ray flux density using the 
unresolved radio hot spot data is possible if we make a few assumptions. 
As discussed by Harris et al. (2000), the ratio of X-ray 
to radio luminosities depends on the ratio $R=u_s/u_B=L_x/L_s$ 
when the X-rays are produced by SSC, where 
$u_s$ is defined above and $u_B= B^2/(8 \pi)$ is the magnetic 
energy density.  The synchrotron photon 
energy density scales as $\theta^{-2}$ (see eq. 1), and 
the magnetic energy density scales as $\theta^{-1.7}$ when 
the magnetic field strength is estimated using minimum energy 
assumptions (Burbidge 1956), 
since $B_{min} \propto \theta^{-6/7}$.  Thus, the 
ratio of energy densities has a very weak dependence on 
$\theta$: $R \propto \theta^{-0.3}$, approximately. 
Since this ratio is so weakly dependent 
on $\theta$, the ratio determined using the unresolved 
flux densities predicts $L_x/L_s$ if the observed 
radio and X-ray emission are produced in the 
same volume.  The X-ray luminosity can be written 
$L_x = \kappa_x L_{\nu_x}{\nu_x}$, and the synchrotron 
luminosity can be written $L_s = \kappa_s L_{\nu_s} \nu_s$. 
Since the X-ray and radio sources are coincident, 
$(L_{\nu_x}\nu_x)/(L_{\nu_s}\nu_s) = (f_{\nu_x}{\nu_x})/ 
(f_{\nu_s}\nu_s)$, where $f_{\nu_{x,s}}$ are the 
observed X-ray and synchrotron flux densities respectively. 
Thus, the ratio R can be computed using $R=u_s/u_B$, and 
the SSC flux density can be predicted using 
$f_{\nu_x} \approx R f_{\nu_s} \nu_s/\nu_x$, 
valid for $\kappa_x \approx \kappa_s$.  Both the 
radio and X-ray flux densities are falling with frequency 
approximately as $\nu^{-1}$, and so both are detected on the 
declining side of the spectrum; in this case it is expected that 
$\kappa_x \approx \kappa_s$. 
 
For the western hot spot region of 3C~254, 
the minimum energy magnetic field 
strength is about 180 $\mu$G, and the synchrotron photon 
energy density is about $6 \times 10^{-12} \hbox{ erg cm}^{-3}$ 
(estimated using the 4.885 GHz data at $0.5\arcsec$ resolution), 
so the ratio of energy densities is about $R \approx 0.005$. 
This ratio predicts 
a 1 keV flux density due to SSC of about 
0.015 nJy, about a factor of 0.025 of the observed flux density. 
If the magnetic field strength is reduced by a factor of 6, 
the observed X-ray flux density matches that observed.  It is 
possible that 
SSC could account for the observed X-ray emission if the 
magnetic field strength is approximately $1/6$ of the minimum 
energy value. 
To reliably compute the predicted 
SSC 1 keV flux density using 
radio data and to solve for the magnetic field strength in 
this region, knowledge of the 
angular diameter of the hot spot is needed in addition to 
the radio spectrum, including the frequencies of 
spectral breaks (that is, a full description of the 
radio spectrum). 
 
For west hot spot region of 3C~280, the minimum energy 
magnetic field is about 
180 $\mu$G and the synchrotron photon energy density is 
about $10^{-11} \hbox{ erg cm}^{-3}$ 
(obtained using the 1.4 GHz data at about $1\arcsec$ resolution), 
so the energy density ratio is $R \approx 0.007$, and 
the predicted 1 keV flux density is about 0.08 nJy. 
This flux density is about 0.1 that observed, implying 
a magnetic field strength of $\sim 1/3$ the 
minimum value, if the X-ray emission 
is produced by SSC.  For the east hot spot region of 3C~280, 
the minimum energy magnetic field is about 170 $\mu$G and 
the synchrotron photon energy density is about 
$4 \times 10^{-12} \hbox{ erg cm}^{-3}$, so the 
ratio R is $R \approx 0.004$, and the predicted 
1 keV flux density is about 0.02 nJy, about 0.06  
of that observed.  The observed X-ray emission 
could be produced by SSC if the magnetic field strength 
is approximately 1/4 of the minimum energy value.

In summary, the X-ray emission detected from the hot spot vicinities 
of 3C~280 and 3C~254 could be produced by SSC if the magnetic 
field strengths are roughly 0.2 to 0.3 of the minimum energy 
values.  Larger magnetic field strengths in and around the 
radio hot spots relative to the minimum 
energy value in that region would produce less X-ray emission 
via SSC. 
However, if the angular sizes of the hot spots are substantially 
smaller than the resolution adopted here, or if the value 
of $\kappa_{\nu}$ is larger than that adopted here, SSC 
could produce the observed X-ray emission from the hot 
spot regions with magnetic field strengths closer to 
the minimum energy values.   
 
\subsection{X-Ray Synchrotron Radiation} 
 
The synchrotron process that produces the radio hot spot 
emission could extend out to X-ray energies and 
produce the X-ray hot spots detected by Chandra. 
This process would require quite recent acceleration, since 
the synchrotron aging timescales for very high energy, relativistic 
electrons that could produce X-rays is quite short 
(on the order of a human lifetime!)  However, 
since acceleration is almost 
certainly occurring in the hot spot region, 
relativistic electrons with very 
high Lorentz factors could conceivably exist there. 
 
For 3C~280, the 1.4 GHz flux density of western 
hot spot is about 2 Jy.  The 1 keV flux density for the western hot 
spot is about 0.79 nJy, which would imply a radio to X-ray 
synchrotron spectral index of about 1.14.  Similarly, the eastern hot 
spot 
1 keV flux density of 0.34 nJy combined with the 1.4 GHz flux 
density of about 0.9 Jy also 
indicates a radio to X-ray 
spectral index of 1.14; these spectral indices are 
consistent with the 1.4, 5, and 
15 GHz radio spectral index determined by Liu, Pooley, \& Riley (1992). 
If the synchrotron radiation extends all the way from radio to X-ray 
energies, the predicted 6200\AA, or 
$5 \times 10^{14}$ Hz, flux density are about 
935 nJy and 420 nJy for the western and eastern hot spots of 3C~280, 
respectively. 
 
\begin{figure} 
\plotone{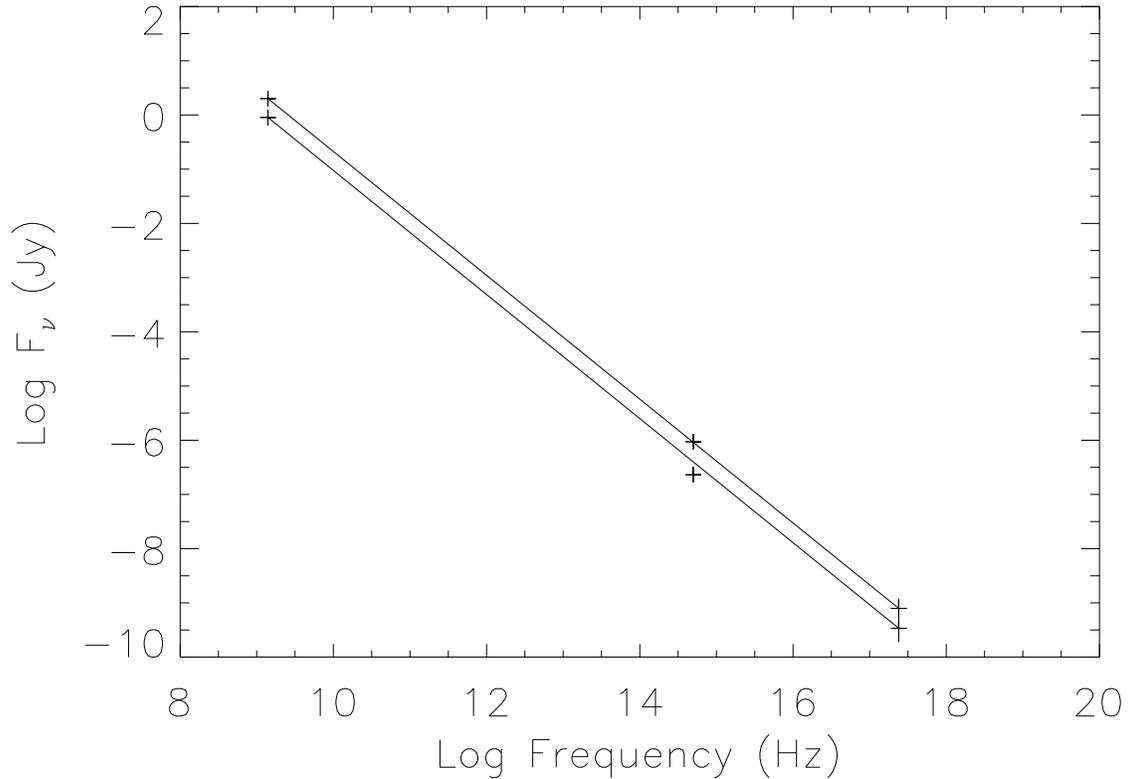} 
\caption{ Radio, optical, and X-ray flux densities for the western (upper  
set of points) and eastern (lower set of points) hot spots in 3C~280. The 
error bars are $3\sigma$. The radio fluxes have uncertainties of 
about 10\%, while the uncertainties in the optical and X-ray flux densities 
include a conservative 5\% flux calibration uncertainty for both 
Chandra and HST. A solid line connects the radio and X-ray points, showing 
that the optical flux density for the  
western hot spot falls just where expected if the radio, optical,  
and X-ray emission are produced by synchrotron radiation.  The 
optical flux density of the   
eastern hot spot is somewhat below 
what one might extrapolate for synchrotron emission, based on the radio 
and X-ray flux densities. Alternatively, one could say the X-ray flux 
density in the eastern hot spot  
is somewhat above what one would expect for synchrotron emission 
based on the radio and optical flux densities. \label{figure:3c280_broadspec}} 
\end{figure}

The hot spot regions of 3C~280 may have optical counterparts 
emitting 6200\AA~ radiation detected by HST (this paper). 
The HST photometry of point sources nearly coincident with 
the radio and X-ray hot spots suggests point 
source flux densities 
about $985 \pm 25$ nJy and 
$230 \pm 15$ nJy for the 
western and eastern hot spots respectively.  These fluxes are certainly 
in the ballpark of the predicted values, and are consistent 
with synchrotron radiation, particularly if the synchrotron spectrum 
has some curvature.  Considering a broken power-law 
with one spectral index $\alpha_{ro}$ between 1.4 GHz and 
6200\AA, and a second spectral index $\alpha_{ox}$ between 
6200\AA~ and 1 keV, these spectral indices that describe the 
data are $\alpha_{ro} \approx 1.13$ and $\alpha_{ox} \approx 
1.15$ for the western hot spot of 3C~280, and $\alpha_{ro} \approx 
1.19$ and $\alpha_{ox} \approx 1.05$ for the eastern hot spot of 
3C~280.  The indices for the western hot spot seem reasonable, 
while those for the eastern hot spot would be quite unusual since 
the high-energy spectral index is expected to be similar to or 
greater than the low-energy spectral index.  Therefore, there may be  
another source of radiation may be contributing to the X-ray emission 
from the eastern hot spot of 3C~280.  If the radio to optical spectral 
index of 1.19 is continued from the optical to the X-ray in the 
eastern hot spot of 3C~280, the 1 keV flux density would be about  
0.15 nJy, about half of the detected X-ray flux density 
(see Figure~\ref{figure:3c280_broadspec}).
 
Thus, the data are consistent with synchrotron radiation producing 
the optical hot spot emission detected by HST. The X-ray emission 
from the western hot spot is consistent with synchrotron radiation, 
but that from the eastern hot spot would require unusual spectral 
properties or an additional source of X-ray emission such as  
SSC or inverse Compton scattering of relativistic electrons with  
CMB photons.   
 
The minimum energy magnetic fields in the hot spot regions 
of 3C~280 
(on scales of about $1\arcsec$) are about 200 $\mu$G. 
Such a magnetic field strength implies 
that the 1 keV emission is produced by relativistic 
electrons with Lorentz factors of about $2 \times 10^7$.  
This conclusion follows from the fact that  
relativistic electrons with Lorentz factors of 
$\gamma$ in a tangled field, for which the field 
component perpendicular  to the electron velocity ($B_{\perp}$) 
satisfies $B_{\perp}= \sqrt{2/3}~B$, produce synchrotron 
radiation with frequency $\nu \approx 30 \gamma^2 B_{-5}$ Hz. Here 
$B_{-5}$ is the magnetic field strength in units of $10^{-5}$G. 
The radiative lifetime of a 
relativistic electron with Lorentz factor of $2 \times 10^7$ in a 
magnetic field of about 200 $\mu$G is only about 30 years 
(see Daly 1992a, eq. 11b).  This estimate assumes a tangled magnetic 
field configuration and that pitch angle 
scattering is important. 
 
For the western hot spot of 3C~254, the 4.885 GHz flux density 
is about 0.1 Jy.  If the 1 keV X-ray emission 
of $\sim 0.5$ nJy 
is synchrotron 
radiation from the same population of relativistic electrons, 
the radio to X-ray spectral index is about 1.09, and the predicted 
6000\AA, or $5 \times 10^{14}$ Hz, flux density is about 450 nJy. 
This is well below the current bounds placed on this emission 
using HST data of $\sim8700$ nJy (Table 2). 
Thus, this X-ray emission could also be produced 
via X-ray synchrotron.  The minimum energy magnetic field on 
a scale of about $1\arcsec$ centered on this hot spot is about 125 $\mu$G, 
implying 1 keV synchrotron radiation would be produced by 
electrons with Lorentz factors of about $2 \times 10^7$, which 
would have radiative lifetimes of about 70 years.

\subsection{Inverse Compton Scattering with the Cosmic Microwave 
Background}

X-ray emission will be produced by inverse Compton scattering 
between relativistic electrons and CMB photons, and with other 
ambient photon fields.  Relativistic electrons 
with Lorentz factors of about $10^3$ will inverse Compton scatter 
CMB photons to an observed energy of about 1 keV,  
independent of any assumptions (including assumptions concerning the 
magnetic field strength) and  
independent of redshift (e.g. Daly 1992a,b).  If the relativistic  
electron energy distribution does not extend below these values 
in the hot spot region, adiabatic expansion will cause a shift 
of the Lorentz factors of the relativistic electrons to  
lower energies in the regions outside of the hot spots, 
as discussed in some detail by Daly (1992b).   
For magnetic field strengths near minimum energy values, 
detected radio emission from outside the hot spots will be  
produced by electrons with Lorentz factors of about  
$10^3$.  Given that the relativistic electron energy 
distribution extends to these Lorentz factors, which it 
almost certainly must, a lower limit can be placed on  
the magnetic field strength of the 
plasma in the immediate vicinity of the radio hot spot by 
calculating the 1 keV X-ray flux density of emission 
produced by inverse Compton scattering with the CMB, and 
requiring that this flux density be less than or equal to the observed 
1 keV flux density.  Using equation (8b) from Daly (1992a), 
and the observed X-ray to radio flux density ratio, the 
magnetic field strength can be determined or bounded.

For the 
western hot spot of 3C~254, the 1 keV to 1.4 GHz flux density ratio 
is about $0.7 \times 10^{-9}$, the radio spectral index is 
about 1.0 (Liu, Pooley, \& Riley 1992), and the 
1.4 GHz radio flux density is about 0.7 Jy. 
Assuming that the X-ray flux produced by IC with the 
CMB is less than or equal to that detected, the magnetic 
field strength in this region must satisfy 
$B \gtrsim 35 \mu$G.  The minimum energy magnetic field 
for the resolved region, obtained using the 1.4 GHz data 
at about $1\arcsec$ resolution and 
adopting the standard assumptions, 
is about 125 $\mu$G.  This result implies that the magnetic field 
strength in the hot spot vicinity 
must be greater than about 0.3 of the minimum 
energy value.    This field strength relative to  
the minimum energy value  
is slightly larger than  
that required for the X-ray emission to be 
completely explained by the SSC process,  
though given how rough the SSC estimate is, we take 
the two constraints to be consistent.   
(It will be improved using higher resolution radio data with 
broad frequency coverage.) 
Similar constraints are obtained using the 
higher resolution 
radio data of Owen \& Puschell (1984) along with the assumption that  
all of the X-ray flux originates from a region that is smaller than 
$0.5\arcsec$. 
 
For the western hot spot region of 3C~280, the 
radio flux density is about 2 Jy at 
1.4 GHz at about $1\arcsec$ resolution; the radio spectral 
index of the western hot spot is roughly 0.8, 
steepening quickly away from the hot spot 
(Liu, Pooley, \& Riley 
1992).  The 1 keV to 1.4 GHz 
flux density ratio is about $4 \times 10^{-10}$. 
Substituting this value into equation (8b) of Daly (1992), 
the magnetic field strength in this 
region satisfies $B \gtrsim 50 \mu$G. 
The minimum energy 
magnetic field strength in this region is about 180 $\mu$G, 
implying that the magnetic field in this region can not 
be less than about 0.3 of the minimum energy value. 
The eastern hot spot has a 1.4 GHz flux density of about 
0.9 Jy, a radio spectral index of about 1, and 
a 1 keV to 1.4 GHz flux density 
ratio of about $4 \times 10^{-10}$, so the 
magnetic field strength in this region must be greater 
than about 65 $\mu$G, or else inverse Compton scattering 
with CMB photons would produce more X-ray emission than is detected. 
The minimum energy magnetic field 
in this region is about 170 $\mu$G, implying that the 
magnetic field strength in the vicinity of the 
radio hot spot 
can not be less than about 
0.4 of the minimum energy value.  Both of these deviations 
from minimum energy conditions are quite similar to and  
consistent with those 
required for SSC to contribute to the detected X-ray 
emission.   
 
Thus, the X-ray and radio flux density ratios  
together with the requirement   
that the inverse Compton scattering between 
relativistic electrons and CMB photons not produce more 
X-rays than observed,  suggest  
that the magnetic field strengths of the hot spot regions can 
not be significantly less than about 0.3 to 0.4 of the minimum energy 
values.  These values are similar to those required if the 
X-ray emission is produced by SSC, though both of these estimates 
are quite rough.  {\em The similarity of these bounds  
suggests that both processes could 
be contributing to the X-ray emission, with SSC being produced 
in the heart of the hot spots, and IC with CMB photons being 
produced in the more extended hot spot vicinity.}   
 
The fact that a lower magnetic field strength  
would over-produce X-rays indicates 
that the relativistic plasma in the  
radio hot spot cannot depart significantly from equipartition  
or minimum energy conditions; the field strength is expected to  
be on the order of or less than the equipartition field strength,  
and the X-ray brightness of the hot spots relative to the radio  
brightness indicates that the field can not be much less than about 
0.3 of the minimum energy value.

The Chandra data also allow an upper bound to be placed on 
X-ray emission from the bridge region of 3C~280 
(see Table~\ref{sources}).  This bound can be compared with 
predicted 1 keV X-ray fluxes produced by inverse 
Compton scattering of microwave background photons 
with the relativistic electron population in the 
bridge region.  Using AIPS\footnote{Astronomical 
Image Processing System (AIPS) is a software package for 
reduction and analysis of radio data.  
http://www.aoc.nrao.edu/aips/}, we estimated  
the 1.4 GHz radio flux density  
within the 150 square arcsecond  
region of the radio bridge that matches that used to obtain the  
X-ray bound to be about 0.4 Jy; this region does {\it not} 
include the radio hot spots.   Most of this flux 
density originates from the extremities of the  
bridge region closest to 
the radio hot spots (though the hot spot regions are  
specifically excluded).  The minimum energy magnetic fields  
in these regions are about 90 $\mu$G for the western bridge 
and about 70 $\mu$G for the eastern bridge (see Table 1 of 
WDW97, converted using $H_0=50$ km s$^{-1}$ Mpc$^{-1}$).  
The radio spectral index in this 
region is about 1 (Liu, Pooley, \& Riley 1992). 
Using eq. (8c) of Daly (1992a), the 1 keV upper bound on 
the X-ray flux  
can be combined with the radio flux density and spectral 
index to determine the smallest value of the magnetic field 
allowed, about 30 $\mu$G.   
Given the 
minimum energy magnetic field strengths listed above, 
this inferred lower limit means that the magnetic field strength 
for the bridge regions near the extremities of the radio bridge  
must be greater than about 0.3 to 0.4 of the minimum energy value 
in this region.  These constraints are similar to those derived 
above, though some care should be taken in interpreting these 
numbers since the X-ray and  
radio flux densities are estimated over a rather large region 
($1\arcsec 
\sim 10 h_{50}^{-1}$ kpc). The  
magnetic field strength over such a large region (150 square 
arcseconds) may 
not be constant.  That is, the radio surface brightness,  
spectral index, and minimum energy magnetic field strength 
vary over this region, so it is almost certainly  
an oversimplification to 
use a total X-ray flux density, total radio flux density, 
single radio spectral index, and single minimum energy  
magnetic field strength to describe this region.    
The constraints obtained are likely to be accurate for 
the extremities of the radio bridge region from which 
most of the radio emission originates, and where most of 
the X-ray emission produced by inverse Compton scattering 
with CMB photons would originate.   
 
The results obtained here, that the bridge magnetic field must  
be greater than about 0.3 of the minimum energy value  
is pushing against the upper bound of about  
0.25 of the minimum energy value  
required  
for consistency between the Chandra bounds on ambient gas densities 
and those predicted by Wellman et al. (1997a) (discussed in  
section 5); given the roughness of the bounds, this 
trend would have to be confirmed with a larger sample of sources, 
or with more Chandra data on these sources.   
It is interesting that the Chandra bounds on the ambient 
gas density tend to require a bridge magnetic field strength that 
is about a factor of 4 below the minimum energy value, while 
the Chandra bounds on inverse Compton scattering from the 
bridge region tend to require a bridge magnetic field strength 
that is on the order of or greater than about 0.3 of the  
minimum energy value.  Since the upper 
and lower bounds are rather close, it will be necessary to consider the 
variations in bridge radio surface brightness, spectral index, 
and magnetic field strength when predicting bridge X-ray emission 
produced via inverse Compton scattering with  
CMB photons.  Similarly, the X-ray data may be of sufficient quality 
to require a more detailed study of the radio predicted ambient 
gas densities.

\section{Summary}

We conclude that not all high redshift  
3CR radio sources are in rich cluster environments with luminous 
intracluster X-ray emission. In particular,  
a cluster like that which surrounds Cygnus A is not present around 
3C~254 or 3C~280.   
Even though the appearance of each radio source is consistent with the 
presence of a high-pressure environment, the luminosity and  
temperature of the surrounding gas could  
be significantly less than that of typical massive clusters of galaxies 
(more like that of a group or a poor cluster). 
These observations indicate that high-redshift radio sources are not 
necessarily the 
signposts of the presence of massive, virialized clusters of galaxies. 
Comparisons of predicted ambient densities from radio observations 
to upper limits of the ambient density of a hot confining medium 
suggest that $b=1/4$ models are yet consistent with the X-ray 
observations, while $b=1$ models are ruled out under most  
assumptions. The sources 3C~254 and 3C~280  
were selected from the WDW97 sample because of  
the confidence in their ROSAT 
properties; coincidentally, their radio-derived ambient densities 
were among the lowest in the WDW97 sample. 
An even more stringent test of the WDW97 predictions could be made with 
Chandra observations of radio sources with higher predicted densities.  
 
The X-ray emission detected from the hot spot regions of 3C~280 
and 3C~254 could be due to SSC emission, X-ray synchrotron emission, 
or inverse Compton scattering between relativistic electrons 
and external photon fields, such as the CMB.  HST, radio, and 
X-ray data from the three hot spots detected by Chandra are 
consistent with X-ray synchrotron emission, though the radiative 
lifetimes of the relativistic electrons producing the X-ray emission 
is likely to be quite short, and would require continuous acceleration 
in the hot spot region.   
SSC production of the X-ray emission can be very roughly estimated, and 
would require magnetic field 
strengths of about 0.2 to 0.3 of the minimum energy values; 
these extremely rough estimates could be significantly improved employing  
high-resolution 
radio observations with broad frequency coverage.   
If SSC dominates, 
the hot spot {\em optical} emisson detected in 3C~280 must have a 
separate origin from the X-ray emission unless the relativistic 
electron energy distribution extends to very low energies. 
Magnetic field  
strengths of 0.3 to 0.4 of the minimum energy values suggests that 
inverse Compton scattering of CMB photons with relativistic electrons 
could also contribute to the total X-ray emission.  As in the 
case of SSC, if IC scattering dominates the X-ray emission,  
the optical and X-ray hot spot emission 
must be produced by different processes unless the relativistic 
electron energy distribution extends to very low energies. 
We cannot conclude from these data  
whether one or more processes are responsible for 
the X-ray emission detected by Chandra and the possible  
optical emission detected by HST from the hot spots of these 
two radio sources. 
 
The comparison of ambient gas densities predicted using radio 
data (WDW97) with the upper 
bounds provided by the Chandra data indicate that the magnetic 
field in the bridge regions of these sources must be below the 
minimum energy values.  The ambient gas densities predicted based on 
radio observations 
are consistent with the bounds placed  
using the Chandra data  
if the 
bridge magnetic field strength is about 1/4 of the minimum energy 
value.  Meanwhile,  
Chandra constraints on inverse Compton scattered X-rays from  
the bridge region tend to push the magnetic field strength in  
this region up relative to the minimum energy value; 
the current bounds are marginally consistent with those obtained 
using the ambient gas density comparisons.   
Here, it is important to keep in mind that the radio 
bridge is a rather large region, and may have a magnetic 
field strength that varies with bridge location.   
Longer Chandra exposures and/or Chandra data 
on similar sources that yield detections rather than bounds  
will be very helpful in determining the bridge magnetic field 
strengths and testing the reliability of the radio determined 
ambient gas densities.  
 
\acknowledgments 
This work was supported by a Chandra X-ray Center 
data analysis grants GO1-2129A and G01-2129B, and by NSF 
grant AST-0096077.   
The Chandra X-ray Observatory Science 
Center (CXC) is operated for NASA by the Smithsonian Astrophysical 
Observatory. 
The HST archival observations are based on observations made with 
the NASA/ESA Hubble Space Telescope, obtained from the data archive 
at the Space Telescope Science Institute. STScI is operated by the 
Association of Universities for Research in Astronomy, Inc. under 
NASA contract NAS5-26555. As a Guest User, we obtained 
additional HST data (WFPC2 associations) 
from the Canadian Astrophysics Data Centre (CADC), which 
is operated by the Herzberg Institute of Astrophysics, National Research 
Council of Canada and from the Space Telescope European Coordinating 
Facilty (ST-ECF), jointly operated by ESA and the European Southern 
Observatory. 
This research has also made use of ROSAT PSPC and HRI 
data obtained from the High Energy Astrophysics Science Archive Research 
Center (HEASARC), provided by NASA's Goddard Space Flight Center. 
We thank Brad Whitmore for useful information regarding 
WFPC2 photometric corrections. 
It is a pleasure to thank Guy Pooley for providing  
radio maps of 3C 254 and 3C 280, and  
Chris O'Dea, Eddie Guerra, 
Jean Eilek, Chris Carilli, Greg Taylor, Rick Perley, 
and Matt Mory for helpful discussions.

\clearpage 
\begin{deluxetable}{lccccc} 
\tablewidth{0pt} 
\tabletypesize{\footnotesize} 
\tablecaption{Compact X-ray Source Results \label{sources}} 
\tablehead{ 
\colhead{ Feature } & \colhead{Count Rate\tablenotemark{e}} & \colhead{Flux (0.5-2.0 
keV)} & \colhead{Power Law Index} 
& \colhead{$f_\nu$ (1 kev)} & \colhead{$L_x$ (2-10 
keV)\tablenotemark{c}} \\ 
\colhead{ }         & \colhead{$10^{-3}$ ct sec$^{-1}$} & 
\colhead{$10^{-15} \flux$} & \colhead{or $kT$ (keV)} 
& \colhead{nJy} & \colhead{$10^{43} \lum$} \\ 
          } 
\startdata 
3c280 nucleus   & $1.94 \pm 0.17$ (96\%) & $3.3 $ & $0.0 \pm 0.36$ & 0.46 & 
$14$ \\ 
3c280 west hot spot & $1.07 \pm 0.13$ (93\%)& $2.4 $ & $2.3 \pm 1.0$  & 0.79 & 
$2.0$ \\ 
                &                                & $2.3 $ & 
$2.2^{+14}_{-1.0}$ keV & &  \\ 
3c280 east hot spot  & $0.60 \pm 0.10$ (89\%) & $1.1 $ & $2.2$\tablenotemark{b} & 
0.34 & $1.2$ \\ 
                &                                & $1.1 $ & 
$2.0$\tablenotemark{b} &  &  \\ 
3c280 bridge\tablenotemark{a}   & $0.55 \pm 0.12$ (67\%)  & $2.1 $ & $2.6 \pm 
1.4$ & $<0.60$ & $<1.5$ \\ 
                &                                & $2.3 $ & 
$1.3^{+0.2}_{-0.6}$ keV & &  \\ 
3c280 bkg sources\tablenotemark{d} & $3.17\pm0.23$ (90\%) & 7.1 & 2.2   &   \\ 
3c254 west hot spot  & $0.84 \pm 0.19$ (81\%) & $1.4 $ & $2.0^{+0.9}_{-0.7}$ & 0.52 
& $0.86$ \\ 
                &                                & $1.9 $ & 
$2.6^{+17}_{-1.5}$ keV &  &  \\ 
3c254 nucleus   & $139 \pm 2$ (100\%)              & $311 $ &  
$1.8\pm0.1$           & 309 & $320$  \\ 
\enddata 
\tablenotetext{a}{Excess counts seem to populate the bridge of 3C~280, 
but their presence is marginally  
statistically consistent with being 
contributions from the hot spots and nucleus, so we conservatively  
treat this measurement as 
an upper limit to the actual bridge X-ray flux.} 
\tablenotetext{b}{Insufficient counts were available for a fit, so a 
power 
law index of 2.2 was assumed to estimate $L_x$.} 
\tablenotetext{c}{Rest frame X-ray luminosity for $H_0=50$ km s$^{-1}$ 
Mpc$^{-1}$; $q_0=0.0$.} 
\tablenotetext{d}{5 compact background sources separated from the 
nucleus by $<120\arcsec$. Sources were extracted and binned into 
a single spectrum to obtain an approximate total count rate and flux.} 
\tablenotetext{e}{Count rates (background-subtracted)  
and background were for 0.3-10.0 keV limits. The rates quoted for 3C~280  
are for the full dataset.   
Background rates were computed locally to the source; the percentage 
of the total counts owing to the source is reported in parentheses.} 
 
\end{deluxetable} 
 
\begin{deluxetable}{lccc} 
\tablewidth{0pt} \tabletypesize{\footnotesize} 
\tablecaption{Summary of HST Observations and Results \label{table:hst}} 
 
\tablehead{ \colhead{Target} & \colhead{3C~254 West Hotspot} & 
\colhead{3C~280 West Hotspot} & \colhead{3C~280 East Hotspot} \\} 
 \startdata 
Chandra Position RA (J2000)   & 11 14 37.7 & 12 56 57.04 & 12 56 58.3 \\ 
 
Chandra Position Dec. (J2000) & 40 37 22.4 & 47 20 19.4  & 47 20 19  \\ 
Radio Position RA (J2000)     & 11 14 37.7 & 12 56 56.93 & 12 56 58.18 
\\ 
Radio Position Dec (J2000)    & 40 37 22.3 & 47 20 19.3  & 47 20 19.2 \\ 
 
HST Position RA (J2000)       & Not det.   & 12 56 56.96 & 12 56 58.26 
\\ 
HST Position Dec. (J2000)     & Not det.   & 47 20 19.9  & 47 20 19.3 
\\ 
HST Program ID                & 5476       &   5401      & 5401   \\ 
Observation Date & Jan 01, 1995 & Aug 22, 1994 & Aug 22, 1994 \\ 
HST Filter (WFPC2 chip) &  F702W (PC)  &    F622W  (WF3)        & F622W 
(WF3)   \\ 
Total HST Exposure (sec) &  $2 \times 140$      & $8 \times 1100$ & $8 \times 
1100$ \\ 
HST Central Wavelength (\AA) & 6918 & 6190 & 6190 \\ 
PHOTFLAM keyword  &  1.88e-18         & 2.81e-18    & 2.81e-18 \\ 
($\, \flambda ({\rm DN \, s}^{-1})^{-1}$) & & & \\ 
Aperture (arcsec) & $1.0\times 1.0$ & $r=0.3$ & $r=0.3$ \\ 
Flux (nJy) & $<8700$\tablenotemark{a} & $985\pm25$ 
  & $230 \pm 15$  \\ 
\enddata 
\tablenotetext{a}{Three-sigma upper limit, uncorrected for CTI or 
geometry ($\leq 10\%$ effects.)} 
\tablenotetext{b}{Uncertainties are one sigma. Fluxes are corrected for 
geometric effects (1\%), CTI (2-3\%), and aperture corrected for 
a $0\farcs3$ aperture to an infinite aperture (16\%).} 
\end{deluxetable} 
 
\begin{deluxetable}{l|rrrrr} 
\tablewidth{0pt} \tabletypesize{\footnotesize} 
\tablecaption{3$\sigma$ upper limits on central surface brightness for 
  29ks image of 3C~254 in the 0.3--2.0 keV band. \label{table:3c254_results}} 
\tablehead{\colhead{} & \multicolumn{5}{c}{$\beta$ value} \\ 
\colhead{ $r_c$ } & \colhead{0.55} & \colhead{0.60} & \colhead{0.65} & 
\colhead{0.70} & \colhead{0.75}\\ 
\colhead{ $h_{50}^{-1}$ kpc} & \multicolumn{5}{c}{counts s$^{-1}$ 
arcmin$^{-2}$} 
} 
\startdata 
 ÊÊÊÊÊÊ150& ÊÊ2.13e-02& Ê2.54e-02& Ê3.00e-02& Ê3.51e-02& Ê4.06e-02 Ê\\ 
 ÊÊÊÊÊÊ200& ÊÊ1.40e-02& Ê1.61e-02& Ê1.84e-02& Ê2.09e-02& Ê2.35e-02 Ê\\ 
 ÊÊÊÊÊÊ250& ÊÊ1.04e-02& Ê1.17e-02& Ê1.31e-02& Ê1.46e-02& Ê1.62e-02 Ê\\ 
 ÊÊÊÊÊÊ300& ÊÊ8.39e-03& Ê9.25e-03& Ê1.02e-02& Ê1.12e-02& Ê1.22e-02 Ê\\ 
 ÊÊÊÊÊÊ350& ÊÊ7.09e-03& Ê7.69e-03& Ê8.36e-03& Ê9.08e-03& Ê9.83e-03 Ê\\ 
\enddata 
\end{deluxetable}

\begin{deluxetable}{l|rrrrr} 
\tablewidth{0pt} \tabletypesize{\footnotesize} 
\tablecaption{3$\sigma$ upper limits on central surface brightness for 
  19ks image of 3C~280 in the 0.3--2.0 keV band. \label{table:3c280_results}} 
\tablehead{\colhead{} & \multicolumn{5}{c}{$\beta$ value} \\ 
\colhead{ $r_c$ } & \colhead{0.55} & \colhead{0.60} & \colhead{0.65} & 
\colhead{0.70} & \colhead{0.75}\\ 
\colhead{ $h_{50}^{-1}$ kpc} & \multicolumn{5}{c}{counts s$^{-1}$ 
arcmin$^{-2}$} 
} 
\startdata 
 ÊÊÊÊÊÊ150& Ê8.09e-03& Ê9.49e-03& Ê1.10e-02& Ê1.26e-02& Ê1.42e-02 \\ 
 ÊÊÊÊÊÊ200& Ê5.44e-03& Ê6.21e-03& Ê7.05e-03& Ê7.92e-03& Ê8.84e-03 \\ 
 ÊÊÊÊÊÊ250& Ê4.09e-03& Ê4.57e-03& Ê5.10e-03& Ê5.66e-03& Ê6.24e-03 \\ 
 ÊÊÊÊÊÊ300& Ê3.29e-03& Ê3.62e-03& Ê3.98e-03& Ê4.36e-03& Ê4.76e-03 \\ 
 ÊÊÊÊÊÊ350& Ê2.78e-03& Ê3.01e-03& Ê3.26e-03& Ê3.53e-03& Ê3.82e-03 \\ 
\enddata 
\end{deluxetable} 
 
\begin{deluxetable}{lccccccc} 
\tablewidth{0pt} \tabletypesize{\footnotesize} 
\tablecaption{Three Sigma Upper Limits on X-ray Luminosities and Electron 
Densities 
\label{table:densities}} 
\tablehead{ \colhead{Name} & 
\colhead{Central rate} & \colhead{$L_{bol}$} & \colhead{$kT_{est}$} & 
\colhead{$n_e$} & \colhead{$r$ } & 
\colhead{$n_e(r)$} & \colhead{$n_{W,pred}$}\\  
\colhead{} & 
\colhead{ct s$^{-1}$ arcmin$^{-2}$} & \colhead{$\lum$} & \colhead{keV} & 
\colhead{$h_{50}^{1/2}$ cm$^{-3}$} 
& \colhead{$h_{50}^{-1}$ kpc} 
& \colhead{$h_{50}^{1/2}$ cm$^{-3}$} &  \colhead{$h_{50}^{20/7}$ 
cm$^{-3}$} } 
\startdata 
3C~254 & 1.46e-2 & 3.4e43 & 1.5  & 9.15e-4 & 78 & 8.3e-4 & $(2.1 \pm 1.0) 
\times 10^{-4}$  \\ 
3C~280 & 5.66e-3 & 3.6e43 & 1.6  & 9.60e-4 & 76 & 8.7e-4 & $(8.3 \pm 2.8) 
\times 10^{-4}$ \\ 
\enddata 
\end{deluxetable} 
 
\begin{deluxetable}{llrrrrrr} 
  \tablewidth{0pt} \tabletypesize{\footnotesize} \tablecaption{Summary 
    of ROSAT Observations\label{table:rosat}} \tablehead{ 
    \colhead{Source} & \colhead{Instr.} & \colhead{Net 
      Counts\tablenotemark{a,b}} & \colhead{Radius} & 
    \colhead{Background\tablenotemark{b}} & \colhead{Exp. Time} & 
    \colhead{Flux (0.5--2.0 keV)} & 
    \colhead{$F_{\nu}$(1 keV) } \\ 
    \colhead{} & \colhead{} & \colhead{} & \colhead{arcsec} & 
    \colhead{cts s$^{-1}$ arcmin$^{-2}$} & \colhead{kilosec} & 
    \colhead{$\flux$} & \colhead{nJy}} \startdata 
  3C280&  PSPC  &  58 $\pm$ 21&  120 & 5.00e-4 & 42.7 & 2.38e-14& 6.6\\ 
  3C280&  HRI   &  41 $\pm$ 40&  120 & 4.44e-3 & 53.3 & 2.22e-14& 6.1\\ 
  3C254&  PSPC  & 713 $\pm$ 29&  150 & 2.30e-4 & 15.6 & 6.12e-13& 168\\ 
  3C254&  HRI   & 971 $\pm$ 35&  150 & 4.93e-3 & 29.2 & 7.29e-13& 201\\ 
  \enddata 
  \tablenotetext{a}{Net counts within a circular aperture with radius 
    given in the Radius column.} 
  \tablenotetext{b}{For the PSPC, the listed values are in 
    the 0.5--2.0 keV band to eliminate contribution from the soft 
    X-ray background, but for the HRI they are in the full 0.1--2.4 
    keV band since the HRI has no spectral resolution.} 
\end{deluxetable}

\clearpage 
 
\end{document}